\documentclass[12pt,a4paper]{article}
\usepackage{graphicx}
\usepackage{amsmath}
\usepackage{amssymb}
\usepackage{bbold}
\usepackage{type1cm}
\usepackage{rotating}
\usepackage{tabularx}
\usepackage{pdflscape}

\usepackage{dsfont}
\usepackage{overpic}
\usepackage{cite}
\usepackage{slashed}
\usepackage{bbm}
\usepackage{subfigure}
\usepackage{graphicx,textcomp}
\allowdisplaybreaks

\author{Peter C. Bruns}
\title{Extended formalism for the study of final-state interaction in $\gamma p\,\rightarrow\,K^{+}\pi\,\Sigma$ photoproduction}
\date{\vspace{-5ex}}

\begin{document}
\maketitle

\begin{center}
Nuclear Physics Institute of the Czech Academy of Sciences, 25068 \v{R}e\v{z}, Czech Republic
\end{center}

\quad \\
\begin{center}
  {\bf Abstract}\\
\end{center}
We extend the formalism for the low-energy analysis of the $\gamma p\rightarrow K^{+}\pi\Sigma$ photoproduction reaction to include the strangeness $S=-1$ meson-baryon p-wave final-state interaction with total angular momentum $J=\frac{3}{2}\,$. As an application, and a check of our method, we evaluate contributions due to the exchange of the $\Sigma^{\ast}(1385)$ resonance as derived from a chiral Lagrangian.  \\

\maketitle

\section{Introduction}

In a series of recent articles \cite{Bruns:2020lyb,Bruns:2022sio,Cieply:2023saa}, we have developed and applied a theoretical framework for the analysis of the lineshape measurements in the $\Lambda(1405)$ resonance region by the CLAS collaboration \cite{CLAS:2013rjt} (see also \cite{Cieply:2022cmu} for some further discussion, and \cite{Nacher:1998mi,Lutz:2004sg,Roca:2013av,Roca:2013cca,Nakamura:2013boa,Mai:2014xna,Wang:2016dtb} for related theoretical work). The lineshapes and $\pi\Sigma$ invariant-mass distributions were determined from the two-meson photoproduction reactions $\gamma p\rightarrow K^{+}\pi\Sigma$ with $\pi\Sigma=\pi^{0}\Sigma^{0},\,\pi^{\pm}\Sigma^{\mp}$, where it was possible to scan through the whole invariant-mass region where the $\Lambda(1405)$ structure is visible. From an initial fit to the data, the authors of \cite{CLAS:2013rjt} incoherently subtracted effects of enhanced final-state interaction (FSI) due to the $K^{\ast}(892),\,\Sigma^{\ast0}(1385),\,\Lambda(1520)$ and $Y^{\ast}(1670)$ to arrive at what they call the residual distribution for the $\Lambda(1405)$. In \cite{Bruns:2022sio}, we compared a set of parameter-free predictions to those results, plugging various different parameterizations for the meson-baryon FSI taken from the literature in our general framework, and observed a strong sensitivity of the predictions to the FSI-model used (we refer also to the thorough comparison of different models in \cite{Cieply:2016jby}). Given that we did not adjust any parameter to fit the mass spectra, the agreement could be judged as reasonable at least for the channels with $\pi^{0}\Sigma^{0}$ and $\pi^{+}\Sigma^{-}$ at the lowest available center-of-mass (c.m.) energy $W\equiv\sqrt{s}=2\,\mathrm{GeV}$. At higher c.m. energies, however, our predictions overshoot the measured spectra. Introducing a form factor for the $K^{+}$ emission in the photoproduction kernel in \cite{Cieply:2023saa}, the latter deficiency could be mended. However, a combined fit of kaon-nucleon scattering data and the mass spectra with 16 adjustable parameters was still found unsatisfactory, especially for the $\pi^{-}\Sigma^{+}$ channel. On the theoretical side, one can think of several reasons for this: First, the photoproduction kernel was essentially constructed from leading-order Baryon Chiral Perturbation Theory, which might simply not be sufficient for an accurate description of the data at the measured energies. Second, focussing on the $\pi\Sigma$ s-wave states, we made several approximations limiting the validity of the description of higher partial waves. And third, the assumption that the $K^{+}$ can be treated as a 'spectator' particle (keeping in mind that effects due to the $\pi K$ resonance $K^{\ast}$ had been subtracted in \cite{CLAS:2013rjt}) might not be well-founded. Obviously, this list of possible shortcomings is not exhaustive. In this contribution, we will address the second point, and extend our framework to include the $\pi\Sigma$ p-wave FSI ($\ell=1,J=1\pm\frac{1}{2}$).  In Sec.~\ref{sec:genformalism}, we shortly recall the general formalism for the two-meson photoproduction process. After that, we discuss the necessary modifications to include the $\pi\Sigma$ p-waves in our framework, which was so far only suited for the treatment of the s-wave FSI. In Sec.~\ref{sec:decuplet}, we apply this formalism to a subset of $\Sigma^{\ast}$ exchange graphs. This is not meant as a full treatment of the decuplet contributions, but only as a consistency check of the modified formalism. We summarize our work in Sec.~\ref{sec:summary}.

\section{General formalism}
\label{sec:genformalism}

The general formalism we recall here was already presented in \cite{Bruns:2020lyb}, and we refer to that reference for a more detailed account. We decompose the invariant amplitude for the process $\gamma(k)p(p_{N})\,\rightarrow\,K^{+}(q_{K})\pi(q_{\pi})\Sigma(p_{\Sigma})$ (where the symbol in brackets denotes the four-momentum of the indicated particle) into Lorentz structures according to
\begin{eqnarray}
\mathcal{M}^{\mu} &=& \gamma^{\mu}\mathcal{M}_{1} + p_{N}^{\mu}\mathcal{M}_{2} + p_{\Sigma}^{\mu}\mathcal{M}_{3} + q_{K}^{\mu}\mathcal{M}_{4} \nonumber \\ &+&  \slashed{k}\left(\gamma^{\mu}\mathcal{M}_{5} + p_{N}^{\mu}\mathcal{M}_{6} + p_{\Sigma}^{\mu}\mathcal{M}_{7} + q_{K}^{\mu}\mathcal{M}_{8}\right) \nonumber \\ &+& \slashed{q}_{K}\left(\gamma^{\mu}\mathcal{M}_{9} + p_{N}^{\mu}\mathcal{M}_{10} + p_{\Sigma}^{\mu}\mathcal{M}_{11} + q_{K}^{\mu}\mathcal{M}_{12}\right) \nonumber \\ &+&  \slashed{q}_{K}\slashed{k}\left(\gamma^{\mu}\mathcal{M}_{13} + p_{N}^{\mu}\mathcal{M}_{14} + p_{\Sigma}^{\mu}\mathcal{M}_{15} + q_{K}^{\mu}\mathcal{M}_{16}\right)\,,\label{eq:MmuDecomp}
\end{eqnarray}
where we omitted structures proportional to the photon four-momentum $k^{\mu}$, $k_{\mu}k^{\mu}\equiv k^2=0$, because they do not contribute to the photoproduction cross section. Gauge invariance of the on-shell amplitude amounts to the requirements
\begin{eqnarray}
  (s-m_{N}^2)\mathcal{M}_{2}\, &\overset{!}{=}& \,(u_{\Sigma}-m_{\Sigma}^2)\mathcal{M}_{3} + (t_{K}-M_{K}^2)\mathcal{M}_{4}\,,\nonumber \\
  2\mathcal{M}_{1}+ (s-m_{N}^2)\mathcal{M}_{6}\, &\overset{!}{=}& \,(u_{\Sigma}-m_{\Sigma}^2)\mathcal{M}_{7} + (t_{K}-M_{K}^2)\mathcal{M}_{8}\,,\nonumber \\
  (s-m_{N}^2)\mathcal{M}_{10}\, &\overset{!}{=}& \,(u_{\Sigma}-m_{\Sigma}^2)\mathcal{M}_{11} + (t_{K}-M_{K}^2)\mathcal{M}_{12}\,,\nonumber \\
  2\mathcal{M}_{9}+ (s-m_{N}^2)\mathcal{M}_{14}\, &\overset{!}{=}& \,(u_{\Sigma}-m_{\Sigma}^2)\mathcal{M}_{15} + (t_{K}-M_{K}^2)\mathcal{M}_{16}\,.\label{eq:gaugeinv}
\end{eqnarray}
These relations feature our basic Mandelstam variables
\begin{eqnarray}
  s &=& (p_{N}+k)^2\,, \qquad u_{\Sigma} = (p_{\Sigma}-k)^2\,, \qquad t_{K} = (q_{K}-k)^2\,, \nonumber \\
  t_{\Sigma} &=& (p_{N}-p_{\Sigma})^2\,, \hspace{-0.25cm}\qquad M_{\pi\Sigma} = \sqrt{(q_{\pi}+p_{\Sigma})^2}\,.\label{eq:mandelstams} 
\end{eqnarray}
We shall make use of two different Lorentz frames: the overall c.m. \hspace{-0.1cm} frame where the nucleon's plus the photon's three-momentum add up to zero, $\vec{p}_{N}+\vec{k}=\vec{0}$, and the so-called $\ast$ frame where $\vec{p}_{\Sigma}^{\,\ast}+\vec{q}_{\pi}^{\,\ast}=\vec{0}$, and where $M_{\pi\Sigma}$ is the c.m. \hspace{-0.1cm}energy of the $\pi\Sigma$ subsystem. Here and in the following, three-momenta, energies and angles referring to the $\ast$ frame will always be labeled by an asterisk. For the convenience of the reader, we give some useful expressions for the energies and momenta in the mentioned reference frames in a table in App.~\ref{app:energies_momenta}. \\

\section{Partial-wave projection}
\label{sec:pwp}

With the help of the general formalism outlined in the previous section, one could now directly start to evaluate Feynman graphs, decompose them into the invariant structures of Eq.~(\ref{eq:MmuDecomp}), and calculate cross sections from them. For the implementation of ``unitarized models'', however, it is of advantage to find those combinations of (partial-wave projections of) the $\mathcal{M}_{i}$ which have simple partial-wave unitarity relations, taking into account the meson-baryon FSI. The (on-shell) amplitudes describing the latter are decomposed into partial waves in a standard fashion,
\begin{eqnarray}
  \mathcal{T}(MB&\rightarrow&\pi\Sigma) = \mathcal{T}^{0} + (\slashed{p}_{\Sigma}+\slashed{q}_{\pi})\mathcal{T}^{1}\,,\label{eq:TpwDecomp} \\
  \mathcal{T}^{0} &=& 4\pi M_{\pi\Sigma}\biggl[\frac{f_{0+}}{\sqrt{E_{\Sigma}^{\ast}+m_{\Sigma}}\sqrt{E_{B}^{\ast}+m_{B}}} - \frac{f_{1-}}{\sqrt{E_{\Sigma}^{\ast}-m_{\Sigma}}\sqrt{E_{B}^{\ast}-m_{B}}} \nonumber\\
    &+& f_{1+}\left(\frac{1}{\sqrt{E_{\Sigma}^{\ast}-m_{\Sigma}}\sqrt{E_{B}^{\ast}-m_{B}}} + \frac{3\cos\alpha^{\ast}}{\sqrt{E_{\Sigma}^{\ast}+m_{\Sigma}}\sqrt{E_{B}^{\ast}+m_{B}}}\right) \nonumber\\
    &-& f_{2-}\left(\frac{1}{\sqrt{E_{\Sigma}^{\ast}+m_{\Sigma}}\sqrt{E_{B}^{\ast}+m_{B}}} + \frac{3\cos\alpha^{\ast}}{\sqrt{E_{\Sigma}^{\ast}-m_{\Sigma}}\sqrt{E_{B}^{\ast}-m_{B}}}\right) + \ldots \biggr]\,,\nonumber\\
 \mathcal{T}^{1} &=& 4\pi\biggl[\frac{f_{0+}}{\sqrt{E_{\Sigma}^{\ast}+m_{\Sigma}}\sqrt{E_{B}^{\ast}+m_{B}}} + \frac{f_{1-}}{\sqrt{E_{\Sigma}^{\ast}-m_{\Sigma}}\sqrt{E_{B}^{\ast}-m_{B}}} \nonumber\\
    &-& f_{1+}\left(\frac{1}{\sqrt{E_{\Sigma}^{\ast}-m_{\Sigma}}\sqrt{E_{B}^{\ast}-m_{B}}} - \frac{3\cos\alpha^{\ast}}{\sqrt{E_{\Sigma}^{\ast}+m_{\Sigma}}\sqrt{E_{B}^{\ast}+m_{B}}}\right) \nonumber\\
    &-& f_{2-}\left(\frac{1}{\sqrt{E_{\Sigma}^{\ast}+m_{\Sigma}}\sqrt{E_{B}^{\ast}+m_{B}}} - \frac{3\cos\alpha^{\ast}}{\sqrt{E_{\Sigma}^{\ast}-m_{\Sigma}}\sqrt{E_{B}^{\ast}-m_{B}}}\right) + \ldots \biggr]\,,\nonumber 
\end{eqnarray} 
for an incoming meson $M$ and baryon $B$ ($\alpha^{\ast}$ is the c.m. scattering angle for this reaction). The $f_{\ell\pm}(M_{\pi\Sigma})$ are the meson-baryon partial-wave scattering amplitudes for $J=\ell\pm\frac{1}{2}$, and the dots stand for terms involving partial waves with $J>3/2$. We are interested here in the $S=-1$ meson-baryon FSI, so the $K^{+}$ emerging from the initial photoproduction reaction is treated as a 'spectator' particle, i.e. its FSI with $M$ and/or $B$ is neglected. This ansatz can obviously only be justified if we limit ourselves to a range of $M_{\pi\Sigma}$ where some $S=-1$ baryon resonance yields the dominant effects of the FSI, or if that channel is known to dominate for other reasons. \\
\begin{figure}[h!]
\centering
\includegraphics[width=0.5\textwidth]{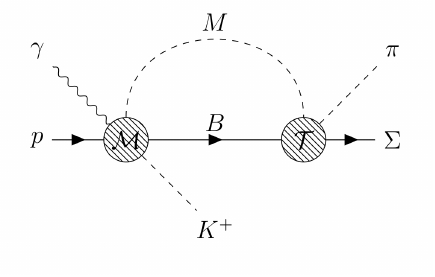}
\caption{Representation of the final-state interaction of the $S=-1$ meson-baryon pair $MB$. $\mathcal{M}$ is the amplitude for $\gamma p\rightarrow K^{+}MB$, and $\mathcal{T}$ is the $S=-1$ meson-baryon scattering amplitude, which can be decomposed into partial waves $f_{\ell\pm}$ as in Eq.~(\ref{eq:TpwDecomp}).}
\label{fig:MBFSI}
\end{figure}%
\\
One can now plug the decomposition of $\mathcal{T}$ and $\mathcal{M}$ in a graph as depicted in Fig.~\ref{fig:MBFSI}, replace the $M$ and $B$ propagators according to the scheme
\begin{equation}
(l^2 - m^2+i\epsilon)^{-1}\,\rightarrow\,(-2\pi i)\Theta(l^{0})\delta(l^2-m^2)\,,
\end{equation}
implied by the Cutkosky rules \cite{Cutkosky:1960sp}, where $\Theta(\cdot)$ denotes the Heaviside step function and $\delta(\cdot)$ the Dirac delta function, evaluate the remaining angular integrals which are not eliminated by the delta functions, and see which parts of the amplitudes $\mathcal{M}_{i}$ ``communicate'' with which $MB$ partial wave $f_{\ell\pm}$ to generate imaginary parts $\mathrm{Im}\,\mathcal{M}_{i}$\,.We decompose the $\mathcal{M}_{i}$ as
\begin{equation}
  \mathcal{M}_{i} = \sqrt{4\pi}\biggl(\mathcal{M}_{i}^{00}\mathcal{Y}_{00} + \sqrt{3}\left(\mathcal{M}_{i}^{10}\mathcal{Y}_{10} + \mathcal{M}_{i}^{11}\left(\frac{\mathcal{Y}_{1,-1}-\mathcal{Y}_{11}}{\sqrt{2}}\right)\right) + \ldots \biggr)\,, \label{eq:newYdecomp}
\end{equation}
where $\mathcal{Y}_{\ell m} \equiv \mathcal{Y}_{\ell m}(\theta^{\ast},\phi^{\ast})$. The dots stand for terms involving spherical harmonics $\mathcal{Y}_{\ell m}$ with $\ell>1$. For the $\mathcal{M}$ plugged in the loop in Fig.~\ref{fig:MBFSI}, $\theta^{\ast},\phi^{\ast}$ are the angles $\theta^{\ast}_{B},\phi^{\ast}_{B}$, defined analogously to the angles $\theta^{\ast}_{\Sigma},\phi^{\ast}_{\Sigma}$, which are fixed while $\theta^{\ast}_{B},\phi^{\ast}_{B}$ are integrated over ($\alpha^{\ast}$ depends on both $\theta^{\ast}_{B},\phi^{\ast}_{B}$ and $\theta^{\ast}_{\Sigma},\phi^{\ast}_{\Sigma}$). Note that there can only be an even dependence on $\phi^{\ast}$, because it enters the Lorentz-invariant amplitudes only through $t_{\Sigma}$ (thus the combination $\mathcal{Y}_{1,-1}-\mathcal{Y}_{11}$). From the orthonormality of the spherical harmonics, we have
\begin{eqnarray}
  \mathcal{M}_{i}^{00} &\equiv& \overline{\mathcal{M}}_{i} := \int\frac{d\Omega^{\ast}_{\Sigma}}{4\pi}\mathcal{M}_{i}(s,M_{\pi\Sigma},t_{K},t_{\Sigma},u_{\Sigma})\,,\label{eq:Mbardef}\\
  \mathcal{M}_{i}^{10} &=& \int\frac{d\Omega^{\ast}_{\Sigma}}{4\pi}\sqrt{\frac{4\pi}{3}}\mathcal{Y}_{10}^{\ast}\,\mathcal{M}_{i}(s,M_{\pi\Sigma},t_{K},t_{\Sigma},u_{\Sigma})\,,\\
  \mathcal{M}_{i}^{11} &=& \int\frac{d\Omega^{\ast}_{\Sigma}}{4\pi}\sqrt{\frac{4\pi}{3}}\frac{(\mathcal{Y}_{1,-1}-\mathcal{Y}_{11})^{\ast}}{\sqrt{2}}\,\mathcal{M}_{i}(s,M_{\pi\Sigma},t_{K},t_{\Sigma},u_{\Sigma})\,,\\
  0 &=& \int\frac{d\Omega^{\ast}_{\Sigma}}{4\pi}\sqrt{\frac{4\pi}{3}}\frac{(\mathcal{Y}_{1,-1}+\mathcal{Y}_{11})^{\ast}}{\sqrt{2}}\,\mathcal{M}_{i}(s,M_{\pi\Sigma},t_{K},t_{\Sigma},u_{\Sigma})\,.
\end{eqnarray}
In our previous publications on the subject \cite{Bruns:2020lyb,Bruns:2022sio,Cieply:2023saa}, we have truncated the expansion $(\ref{eq:newYdecomp})$ after the first term, and taken into account only the $\overline{\mathcal{M}}_{i}$ (see e.g. the remark above Eq.~(2.8) in \cite{Bruns:2022sio}). This was sufficient for studying the effect of s-wave resonance poles in the FSI, but it won't suffice for baryon resonances with $J=3/2,\,\ell=1$. In fact, in Sec.~\ref{sec:decuplet} and App.~\ref{app:Mi_decuplet} below, we shall evaluate such pole terms, and find that the corresponding terms in $\mathcal{M}_{3,7,11,15}$ are independent of the baryon angles, while the remaining $\mathcal{M}_{i}$ depend linearly on $t_{\Sigma}$ and $u_{\Sigma}$ (which can be written as linear combinations of $\mathcal{Y}_{00}(\theta^{\ast},\phi^{\ast})$, $\mathcal{Y}_{10}(\theta^{\ast},\phi^{\ast})$ and $\mathcal{Y}_{1,-1}(\theta^{\ast},\phi^{\ast})-\mathcal{Y}_{11}(\theta^{\ast},\phi^{\ast})$). Thus, our ansatz will be to take into account $12+8+8=28$ independent amplitude projections,
\begin{equation}\label{list:Mellemmi}
  \mathcal{M}^{00}_{1,2,3,5,6,7,9,10,11,13,14,15}\,, \quad \mathcal{M}^{10}_{1,2,5,6,9,10,13,14}\,, \quad \mathcal{M}^{11}_{1,2,5,6,9,10,13,14}\,.
\end{equation}
The amplitude projections with $i=4,8,12,16$ can be fixed from the ones above with the help of the gauge invariance constraints (\ref{eq:gaugeinv}). From the set (\ref{list:Mellemmi}), we can construct four amplitudes $\mathcal{C}^{j}_{0+}$ with $\ell\pm=0+$, four amplitudes $\mathcal{C}^{j}_{1-}$ with $\ell\pm=1-$, ten amplitudes $\mathcal{C}^{j}_{1+}$ and ten amplitudes $\mathcal{C}^{j}_{2-}$, $4+4+10+10=28$. The relations can be inverted, so the $\mathcal{M}^{lm}_{i}$ of (\ref{list:Mellemmi}) can be expressed through the 28 partial-wave projections $\mathcal{C}^{j}_{\ell\pm}$ just mentioned. We point out that, if we {\em require}\, that the partial-wave projections $\mathcal{C}^{j}_{1+}=\mathcal{C}^{j}_{2-}=0$, this procedure leads us exactly to the form of the amplitude given by the $\mathcal{M}_{i}^{\mathcal{C}}$ in App.~D of \cite{Bruns:2020lyb} (see also (A.10)-(A.21) in \cite{Bruns:2022sio} for the case where also $\mathcal{C}^{j}_{1-}=0$), which demonstrates the consistency with our earlier work.\\
\quad\\
The $0+,1-$ partial-wave projections in our improved formalism are given by
\begin{eqnarray}
\mathcal{C}^{j}_{\ell\pm} &=& \mathcal{C}^{j[1]}_{\ell\pm} + \Delta\mathcal{C}^{j}_{\ell\pm}\,,\\
\Delta\mathcal{C}^{j}_{\ell\pm} &=& \frac{|\vec{p}^{\,\ast}_{\Sigma}|}{(E_{\Sigma}^{\ast} \pm m_{\Sigma})E_{\gamma}^{\ast}}\biggl(\sum_{k=1}^{16}c^{j}_{\ell\pm}(k)\mathcal{M}_{k}^{10}+\frac{\sum_{k=1}^{16}d^{j}_{\ell\pm}(k)\mathcal{M}_{k}^{11}}{(s-m_{N}^2)|\vec{q}_{K}|\sin\theta_{K}}\biggr)\,.\label{eq:DeltaC}
\end{eqnarray}
Note that $|\vec{q}_{K}|\sin\theta_{K}$ relates to the overall c.m. frame, {\em not\,} the $\ast$ frame (we refer again to App.~\ref{app:energies_momenta} for the definition of the energies, momenta and angles used here). 
The $\mathcal{C}^{j[1]}_{\ell\pm}$ are given by combinations of the $\mathcal{M}_{i}^{00}$, and equal the expressions found earlier in \cite{Bruns:2020lyb}. For easy reference, we reproduce them here in App.~\ref{app:old_Cs}. The non-vanishing coefficients $c^{j}_{\ell\pm}(k)$ and $d^{j}_{\ell\pm}(k)$ are given in App.~\ref{app:cd_coefficients}. All $c^{j}_{\ell\pm}(k)\,,\,d^{j}_{\ell\pm}(k)\,$ not listed there are to be set to zero. - We define furthermore
\begin{eqnarray*}
  \mathcal{A}_{0+}^{1} &:=& \sqrt{E_{\Sigma}^{\ast}+m_{\Sigma}}\sqrt{E_{\pi\Sigma}+M_{\pi\Sigma}}\,\left(\mathcal{C}_{0+}^{1}\right)\sqrt{E_{N}-m_{N}}\,/\sqrt{2M_{\pi\Sigma}}\,,\\
  \mathcal{A}_{0+}^{2} &:=& \sqrt{E_{\Sigma}^{\ast}+m_{\Sigma}}\sqrt{E_{\pi\Sigma}-M_{\pi\Sigma}}\,\left(\mathcal{C}_{0+}^{2}\right)\sqrt{E_{N}+m_{N}}\,/\sqrt{2M_{\pi\Sigma}}\,,\\
  \mathcal{A}_{0+}^{3} &:=& \sqrt{E_{\Sigma}^{\ast}+m_{\Sigma}}\sqrt{E_{\pi\Sigma}+M_{\pi\Sigma}}\,\left(\mathcal{C}_{0+}^{3}\right)\sqrt{E_{N}-m_{N}}\,/\sqrt{2M_{\pi\Sigma}}\,,\\
  \mathcal{A}_{0+}^{4} &:=& \sqrt{E_{\Sigma}^{\ast}+m_{\Sigma}}\sqrt{E_{\pi\Sigma}-M_{\pi\Sigma}}\,\left(\mathcal{C}_{0+}^{4}\right)\sqrt{E_{N}+m_{N}}\,/\sqrt{2M_{\pi\Sigma}}\,,\\
  \mathcal{A}_{1-}^{1} &:=& \sqrt{E_{\Sigma}^{\ast}-m_{\Sigma}}\sqrt{E_{\pi\Sigma}+M_{\pi\Sigma}}\,\left(\mathcal{C}_{1-}^{1}\right)\sqrt{E_{N}+m_{N}}\,/\sqrt{2M_{\pi\Sigma}}\,,\\
  \mathcal{A}_{1-}^{2} &:=& \sqrt{E_{\Sigma}^{\ast}-m_{\Sigma}}\sqrt{E_{\pi\Sigma}-M_{\pi\Sigma}}\,\left(\mathcal{C}_{1-}^{2}\right)\sqrt{E_{N}-m_{N}}\,/\sqrt{2M_{\pi\Sigma}}\,,\\
  \mathcal{A}_{1-}^{3} &:=& \sqrt{E_{\Sigma}^{\ast}-m_{\Sigma}}\sqrt{E_{\pi\Sigma}+M_{\pi\Sigma}}\,\left(\mathcal{C}_{1-}^{3}\right)\sqrt{E_{N}+m_{N}}\,/\sqrt{2M_{\pi\Sigma}}\,,\\
  \mathcal{A}_{1-}^{4} &:=& \sqrt{E_{\Sigma}^{\ast}-m_{\Sigma}}\sqrt{E_{\pi\Sigma}-M_{\pi\Sigma}}\,\left(\mathcal{C}_{1-}^{4}\right)\sqrt{E_{N}-m_{N}}\,/\sqrt{2M_{\pi\Sigma}}\,.
\end{eqnarray*}
In case one deals with more $MB$ channels than just $\pi\Sigma$, one can form a vector of each of these $\mathcal{A}_{\ell\pm}^{j}$, one component for each $MB$ channel replacing $\pi\Sigma$. Likewise, the meson-baryon partial waves $f_{\ell\pm}$ are then considered as matrices acting in this ``channel space'' (one component for each reaction $MB\rightarrow M'B'$), and the partial-wave unitarity relations we find from the analysis of the graph of Fig.~\ref{fig:MBFSI} with the help of the Cutkosky rules can be written in the form of simple matrix equations,
\begin{equation}\label{eq:uniA}
\mathrm{Im}(\mathcal{A}^{i}_{\ell\pm}) = (f_{\ell\pm})^{\dagger}(|\vec{p}^{\,\ast}|)(\mathcal{A}^{i}_{\ell\pm})\,,
\end{equation}
where the dagger denotes hermitian conjugation of the channel matrix, and $(|\vec{p}^{\,\ast}|)$ is a diagonal channel matrix with entries of baryon momenta $|\vec{p}^{\,\ast}_{B}|$ for each $MB$ channel, defined just like $|\vec{p}^{\,\ast}_{\Sigma}|$ in App.~\ref{app:energies_momenta}. The partial-wave cross sections due to the $\pi\Sigma$ $\ell\pm=0+,1-$ projected amplitudes can be conveniently expressed through the $\mathcal{A}^{i}_{\ell\pm}$, see Eq.~(27) in \cite{Bruns:2020lyb}, or Eq.~(2.9) in \cite{Bruns:2022sio} for the $\pi\Sigma$ s-wave. We would like to stress that these partial-wave cross section formulas can be derived from a {\em gauge-invariant\,} set of amplitudes like the $\mathcal{M}_{i}^{\mathcal{C}}$ from App.~D of \cite{Bruns:2020lyb}, mentioned above, where $\mathcal{M}_{4,8,12,16}^{\mathcal{C}}$ are expressed through the remaining $\mathcal{M}_{i}^{\mathcal{C}}$'s via the gauge-invariance constraints (\ref{eq:gaugeinv}), so there can be no conflict between our results for the partial-wave cross sections and gauge invariance. This can be done for {\em any\,} given (e.g., unitarized) model for the $\mathcal{A}^{i}_{\ell\pm}$. \\

\newpage

The results above were already presented in \cite{Bruns:2020lyb,Bruns:2022sio,Cieply:2023saa}, except for the modification due to the $\Delta\mathcal{C}^{j}_{\ell\pm}$. We now turn to the higher partial waves. The linear combinations projecting on the $1+,2-$ $MB$ partial waves can be taken as  
\begin{eqnarray*}
  \mathcal{C}_{1+}^{1} &=& \mathcal{M}_{3}^{00}-(M_{\pi\Sigma}-m_{N})\mathcal{M}_{11}^{00} + (s-m_{N}^2)\mathcal{M}_{15}^{00}\,,\\
  \mathcal{C}_{1+}^{2} &=& \mathcal{M}_{7}^{00}+\mathcal{M}_{11}^{00} - (M_{\pi\Sigma}+m_{N})\mathcal{M}_{15}^{00}\,,\\
  \mathcal{C}_{1+}^{3} &=& \mathcal{M}_{1}^{10} -(M_{\pi\Sigma}+m_{N})\mathcal{M}_{9}^{10} + (s-m_{N}^2)\mathcal{M}_{13}^{10}\,,\\
  \mathcal{C}_{1+}^{4} &=& \mathcal{M}_{1}^{11} -(M_{\pi\Sigma}+m_{N})\mathcal{M}_{9}^{11} + (s-m_{N}^2)\mathcal{M}_{13}^{11}\,,\\
  \mathcal{C}_{1+}^{5} &=& \mathcal{M}_{2}^{10} + 2\mathcal{M}_{9}^{10} - (M_{\pi\Sigma}-m_{N})\mathcal{M}_{10}^{10} + (s-m_{N}^2)\mathcal{M}_{14}^{10}\,,\\
  \mathcal{C}_{1+}^{6} &=& \mathcal{M}_{2}^{11} + 2\mathcal{M}_{9}^{11} - (M_{\pi\Sigma}-m_{N})\mathcal{M}_{10}^{11} + (s-m_{N}^2)\mathcal{M}_{14}^{11}\,,\\
  \mathcal{C}_{1+}^{7} &=& \mathcal{M}_{5}^{10} + \mathcal{M}_{9}^{10} -(M_{\pi\Sigma}-m_{N})\mathcal{M}_{13}^{10}\,,\\
  \mathcal{C}_{1+}^{8} &=& \mathcal{M}_{5}^{11} + \mathcal{M}_{9}^{11} -(M_{\pi\Sigma}-m_{N})\mathcal{M}_{13}^{11}\,,\\
  \mathcal{C}_{1+}^{9} &=& \mathcal{M}_{6}^{10} + \mathcal{M}_{10}^{10} -2\mathcal{M}_{13}^{10} -(M_{\pi\Sigma}+m_{N})\mathcal{M}_{14}^{10}\,,\\
  \mathcal{C}_{1+}^{10} &=& \mathcal{M}_{6}^{11} + \mathcal{M}_{10}^{11} -2\mathcal{M}_{13}^{11} -(M_{\pi\Sigma}+m_{N})\mathcal{M}_{14}^{11}\,,\\
  \quad & & \quad \\
  \mathcal{C}_{2-}^{1} &=& \mathcal{M}_{3}^{00}+(M_{\pi\Sigma}+m_{N})\mathcal{M}_{11}^{00} + (s-m_{N}^2)\mathcal{M}_{15}^{00}\,,\\
  \mathcal{C}_{2-}^{2} &=& \mathcal{M}_{7}^{00}+\mathcal{M}_{11}^{00} + (M_{\pi\Sigma}-m_{N})\mathcal{M}_{15}^{00}\,,\\
  \mathcal{C}_{2-}^{3} &=& \mathcal{M}_{1}^{10} +(M_{\pi\Sigma}-m_{N})\mathcal{M}_{9}^{10} + (s-m_{N}^2)\mathcal{M}_{13}^{10}\,,\\
  \mathcal{C}_{2-}^{4} &=& \mathcal{M}_{1}^{11} +(M_{\pi\Sigma}-m_{N})\mathcal{M}_{9}^{11} + (s-m_{N}^2)\mathcal{M}_{13}^{11}\,,\\
  \mathcal{C}_{2-}^{5} &=& \mathcal{M}_{2}^{10} + 2\mathcal{M}_{9}^{10} + (M_{\pi\Sigma}+m_{N})\mathcal{M}_{10}^{10} + (s-m_{N}^2)\mathcal{M}_{14}^{10}\,,\\
  \mathcal{C}_{2-}^{6} &=& \mathcal{M}_{2}^{11} + 2\mathcal{M}_{9}^{11} + (M_{\pi\Sigma}+m_{N})\mathcal{M}_{10}^{11} + (s-m_{N}^2)\mathcal{M}_{14}^{11}\,,\\
  \mathcal{C}_{2-}^{7} &=& \mathcal{M}_{5}^{10} + \mathcal{M}_{9}^{10} +(M_{\pi\Sigma}+m_{N})\mathcal{M}_{13}^{10}\,,\\
  \mathcal{C}_{2-}^{8} &=& \mathcal{M}_{5}^{11} + \mathcal{M}_{9}^{11} +(M_{\pi\Sigma}+m_{N})\mathcal{M}_{13}^{11}\,,\\
  \mathcal{C}_{2-}^{9} &=& \mathcal{M}_{6}^{10} + \mathcal{M}_{10}^{10} -2\mathcal{M}_{13}^{10} +(M_{\pi\Sigma}-m_{N})\mathcal{M}_{14}^{10}\,,\\
  \mathcal{C}_{2-}^{10} &=& \mathcal{M}_{6}^{11} + \mathcal{M}_{10}^{11} -2\mathcal{M}_{13}^{11} +(M_{\pi\Sigma}-m_{N})\mathcal{M}_{14}^{11}\,.
\end{eqnarray*}
The unitarity relations (analogous to the ones for the $\mathcal{A}^{i}_{0+,1-}$) can be satisfied by
\begin{displaymath}
\sqrt{E_{\Sigma}^{\ast}\pm m_{\Sigma}}\,|\vec{p}_{\Sigma}^{\,\ast}|\,\mathcal{C}_{\ell\pm}^{i}\quad\mathrm{for}\quad i=1,2\,,\qquad \sqrt{E_{\Sigma}^{\ast}\pm m_{\Sigma}}\,\mathcal{C}_{\ell\pm}^{i}\quad\mathrm{for}\quad i=3,\ldots,10\,.
\end{displaymath}
As before, $\pi$ and $\Sigma$ (their masses, couplings,\,\ldots) are to be replaced by $M$ and $B$ if the projected amplitude for the intermediate process $\,\gamma p\,\rightarrow\,K^{+}MB\,$ (see Fig.~\ref{fig:MBFSI}) is considered. In the following we will disregard the $2-$ contribution, as well as the $\pi\Sigma$ partial waves with $J>3/2$.  The double-differential cross section is then given by
\begin{equation}
 \frac{d^2\sigma}{d\Omega_{K}dM_{\pi\Sigma}} = \frac{|\vec{q}_{K}||\vec{p}_{\Sigma}^{\,\ast}|}{(4\pi)^4s|\vec{k}|}\biggl[|\mathcal{A}|^2_{0+} + |\mathcal{A}|^2_{1-} + |\mathcal{A}|^2_{1+}\biggr]\,,\label{eq:d2csA}
\end{equation}
where the first two terms refer to the expressions given in \cite{Bruns:2020lyb,Bruns:2022sio,Cieply:2023saa}, and the (very lengthy) expression for $|\mathcal{A}|^2_{1+}$, involving the $\mathcal{C}_{1+}^{i}$, is given in App.~\ref{app:Asqr1plus}. We have verified explicitly that there are no interference terms between the different partial waves ($0+,1-,1+$) in (\ref{eq:d2csA}), providing yet another check for the consistency of our formalism. The remarks on gauge invariance made above apply analogously also for this more general formula involving the $1+$ partial wave.

\clearpage

\section{Decuplet contributions: $\Sigma^{\ast 0}$ exchange}
\label{sec:decuplet}

We shall use a conventional Lagrangian for the decuplet-$MB$ interaction (see e.g. \cite{Jenkins:1991ne,Jenkins:1992pi,Lebed:1993yu}):
\begin{equation}\label{eq:LDMB}
  \mathcal{L}_{\mathrm{DMB}} = -\frac{\mathcal{C}}{2}\epsilon^{ilm}(\bar{T}^{\mu})_{ijk}(u_{\mu})^{j}_{\,\,l}B^{k}_{\,\,m} \,+\,\mathrm{h.c.}\,,
\end{equation}
where the Rarita-Schwinger field $T$ contains the decuplet fields and is totally symmetric in the flavor indices $i,j,k\ldots$ (for example, $T_{111}=\Delta^{++}$, $T_{123}=\Sigma^{\ast 0}/\sqrt{6}$),
the octet baryon and pseudo-Goldstone-boson fields are collected in matrices $B$ and $\Phi$, respectively,
\begin{equation*} \label{eq:B}
B = \left(
\begin{array}{ccc}
\frac{1}{\sqrt{2}}\Sigma^{0}+\frac{1}{\sqrt{6}}\Lambda & \Sigma^{+} & p \\
\Sigma^{-} & -\frac{1}{\sqrt{2}}\Sigma^{0}+\frac{1}{\sqrt{6}}\Lambda & n \\
\Xi^{-} & \Xi^{0} & -\frac{2}{\sqrt{6}}\Lambda \end{array}\right)\,,
\end{equation*}
\begin{equation*}\label{eq:Phi}
\Phi = \left(
\begin{array}{ccc}
\frac{1}{\sqrt{2}}\pi^{0}+\frac{1}{\sqrt{6}}\eta_{8} & \pi^{+} & K^{+} \\
\pi^{-} & -\frac{1}{\sqrt{2}}\pi^{0}+\frac{1}{\sqrt{6}}\eta_{8} & K^{0} \\
K^{-} & \bar K^{0} & -\frac{2}{\sqrt{6}}\eta_{8} \end{array}\right),
\end{equation*}
$U=u^2=\mathrm{exp}\left(i\sqrt{2}\Phi/F_{0}\right)$, $\nabla_{\mu}U=\partial_{\mu}U-i\lbrack v_{\mu},\,U\rbrack$, $u_{\mu}=iu^{\dagger}(\nabla_{\mu}U)u^{\dagger}$, $v_{\mu}=-eQA_{\mu}$ contains the electromagnetic four-potential and the quark charge matrix $Q$, $F_{0}$ is the meson decay constant in the three-flavor chiral limit, $\epsilon^{\cdots}$ denotes the Levi-Civita symbol, and roughly $|\mathcal{C}|\sim 3/2\,$. \\
The vertex rules for $T^{\nu}M(q)B$ we need in the following are of the simple form
\begin{displaymath}
\frac{\nu_{MB}\mathcal{C}}{2\sqrt{3}F_{0}}q^{\nu}\,,
\end{displaymath}
where $\nu_{MB}=1$ for $p\rightarrow K^{+}\Sigma^{\ast 0}$ and $\nu_{MB}=0,-1,1$ for $\Sigma^{\ast 0}\rightarrow \pi^{0}\Sigma^{0},\,\pi^{-}\Sigma^{+},\,\pi^{+}\Sigma^{-}$, respectively. \\
We also need the $\gamma TMB$ vertex following from minimal coupling (for a meson charge $Q_{M}$),
\begin{displaymath}
-eQ_{M}\frac{\nu_{MB}\mathcal{C}}{2\sqrt{3}F_{0}}g^{\mu\nu}\,,
\end{displaymath}
and the spin $3/2$ propagator in momentum-space (in $d=4$ dimensions),
\begin{equation}\label{eq:RSprop}
-\frac{i(\slashed{p}+m_{D})}{p^2-m_{D}^2+i\epsilon}\biggl[g^{\mu\nu}-\frac{1}{3}\gamma^{\mu}\gamma^{\nu}-\frac{\gamma^{\mu}p^{\nu}-\gamma^{\nu}p^{\mu}}{3m_{D}}-\frac{2}{3}\frac{p^{\mu}p^{\nu}}{m_{D}^2}\biggr]\,.
\end{equation}
Other choices for the interaction in Eq.~(\ref{eq:LDMB}) and the propagator in Eq.~(\ref{eq:RSprop}) can be found in the literature, but the according results differ just in a set of contact terms (compare e.g. the discussion in \cite{Krebs:2009bf}). As we are mainly interested in the effect due to the pole terms, the above formalism is sufficient for our purposes. In particular, as we work at a fixed perturbative order in an EFT framework, the famous consistency problems \cite{Johnson:1960vt,Velo:1969bt} afflicting the theory of higher-spin particles minimally coupled to an electromagnetic field will not concern us here. For a more realistic description of the hyperon exchange effects in the photoproduction process, one would of course have to go beyond the minimal-coupling scheme used in this work. The vertices involving only the octet baryons and/or mesons are derived from the standard leading-order chiral Lagrangians, see e.g. Sec.~3 of \cite{Bruns:2020lyb}.\\
\clearpage
\begin{figure}[h!]
\centering
\includegraphics[width=0.75\textwidth]{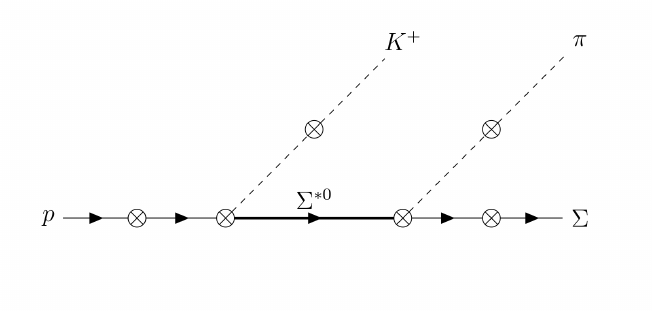}
\caption{Exchange graphs of $\Sigma^{\ast 0}$ with minimal coupling of the photon. The figure represents six different graphs, in each of which the photon couples to one of the crossed vertices.}
\label{fig:Sigstar0exchange}
\end{figure}%
The six graphs we consider here are represented in Fig.~\ref{fig:Sigstar0exchange}, and the corresponding contributions to the invariant amplitudes $\mathcal{M}_{i}$ are given in App.~\ref{app:Mi_decuplet}. Only the graphs where the photon couples to one of the first three (leftmost) vertices contain a pole at $M_{\pi\Sigma}=m_{D}$ (the mass of the decuplet member), while the pole in the remaining three contributions, where the photon couples to a vertex to the right of the $\Sigma^{\ast 0}$ propagator, is outside of the physical region for the photoproduction process. The ``genuine pole terms'' are given by the contributions to the $\mathcal{M}_{i}$ due to the first three graphs, labeled by a superscript $(a)$ in App.~\ref{app:Mi_decuplet}, with $M_{\pi\Sigma}$ set to $m_{D}$ everywhere except in the denominator $\sim m_{D}^2-M_{\pi\Sigma}^2\,$. Since the exchanged hyperon is neutral, the first and the last three graphs form two gauge-invariant subsets of amplitudes.\\

If we consider only the terms with a pole at $M_{\pi\Sigma}=m_{D}$ (the ``genuine pole terms'', which are gauge invariant for themselves), contained in the $\mathcal{M}_{i}^{(a)}$, we get a pure $\ell\pm=1+$ amplitude, with $\mathcal{C}_{0+}^{i}=0\,$, $\mathcal{C}_{1-}^{i}=0\,$, $i=1,\ldots,4$, and also $\mathcal{C}_{2-}^{j}=0$, $j=1,\ldots,10$, while 
\begin{eqnarray}
  \mathcal{C}_{1+}^{1} &=& 6m_{D}n_{D}(M_{\pi\Sigma})\,,\quad \mathcal{C}_{1+}^{2,3,4} = 0\,,\nonumber\\
  \mathcal{C}_{1+}^{5} &=& 2n_{D}(M_{\pi\Sigma})|\vec{p}_{\Sigma}^{\,\ast}|\left(\frac{s+m_{D}^2-M_{K}^2}{s-m_{N}^2}-\frac{m_{D}}{E_{\gamma}^{\ast}}\right)\,,\nonumber\\
  \mathcal{C}_{1+}^{6} &=& 2n_{D}(M_{\pi\Sigma})|\vec{p}_{\Sigma}^{\,\ast}|\frac{|\vec{q}_{K}|}{E_{\gamma}^{\ast}}\sin\theta_{K}\,,\nonumber\\
  \mathcal{C}_{1+}^{7} &=& \frac{1}{2}\mathcal{C}_{1+}^{5}\,,\quad \mathcal{C}_{1+}^{8} = \frac{1}{2}\mathcal{C}_{1+}^{6}\,,\quad \mathcal{C}_{1+}^{9,10} = 0\,,\label{eq:C1p_poleapprox}\\
  n_{D}(M_{\pi\Sigma}) &:=& \frac{e\,\mathcal{C}^2}{36F_{\pi}F_{K}}\frac{\nu_{\pi\Sigma}}{m_{D}^2-M_{\pi\Sigma}^2}\,,\nonumber
\end{eqnarray}
where we set $M_{\pi\Sigma}=m_{D}$ everywhere except in $n_{D}(M_{\pi\Sigma})$, and replaced $F_{0}$ by $F_{\pi}$ ($F_{K}$) at the vertex where a pion (kaon) is emitted (the difference is of higher chiral order). Recall that $|\nu_{\pi\Sigma}|=1$ for the $\pi^{\pm}\Sigma^{\mp}$ cases. The fact that the pole at $M_{\pi\Sigma}=m_{D}$ appears only in the $\ell=1,\,J=\frac{3}{2}$ amplitudes is an excellent check for our partial-wave projection formulas. \\

\newpage

\begin{figure}[!h]
\centering
\subfigure[\,$\frac{d\sigma}{dM_{\pi\Sigma}}\bigl|_{\mathrm{pole}}$]{\includegraphics[width=0.49\textwidth]{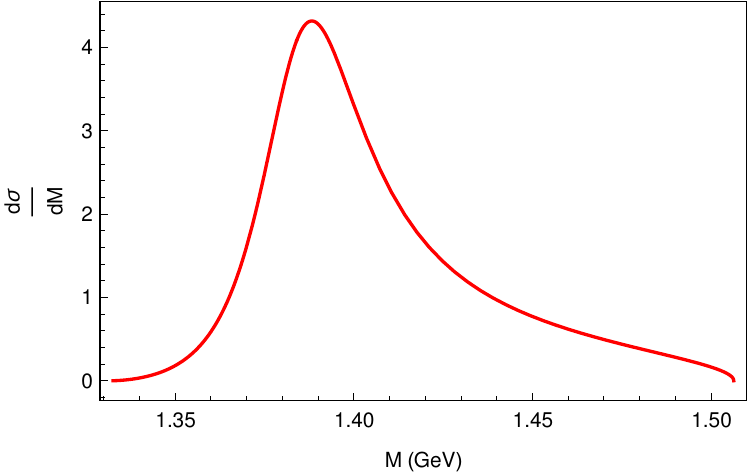}}
\subfigure[\,$\frac{d\sigma}{dz_{K}}\bigl|_{\mathrm{pole}}$]{\includegraphics[width=0.49\textwidth]{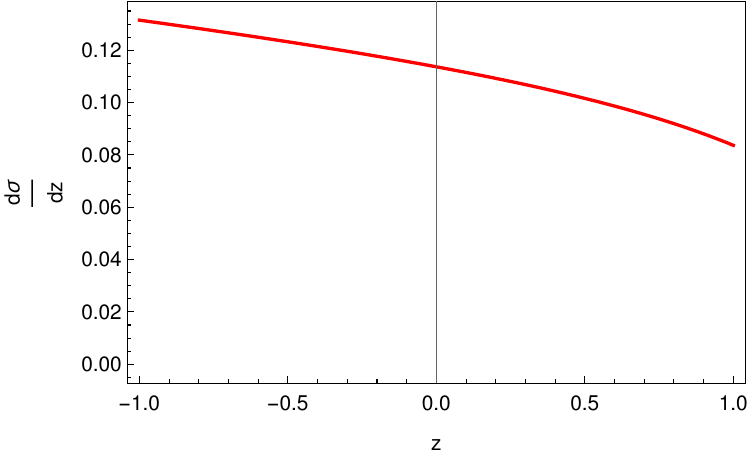}}
\caption{$\frac{d\sigma}{dM_{\pi\Sigma}}$ and $\frac{d\sigma}{dz_{K}}$ from $\Sigma^{\ast}$ pole terms for $\gamma p\rightarrow K^{+}\pi^{\mp}\Sigma^{\pm}$, $\sqrt{s}=2\,\mathrm{GeV}$, in $\mu$b/GeV and $\mu$b, respectively. The pole terms are the same for $\pi^{-}\Sigma^{+}$ and $\pi^{+}\Sigma^{-}$, except for the different $\Sigma$ masses, which we take here as the average, $m_{\Sigma}=\frac{1}{2}(m_{\Sigma^{+}}+m_{\Sigma^{-}})$. Furthermore we set $m_{D}=m_{\Sigma^{\ast 0}}=1.384\,\mathrm{GeV}$, $\Gamma_{\Sigma^{\ast 0}}=36\,\mathrm{MeV}$, $|\mathcal{C}|=\frac{3}{2}$, $F_{\pi}=92.2\,\mathrm{MeV}$, $F_{K}=110.1\,\mathrm{MeV}$.}
\label{fig:Sigmastarexchange_poleterms}
\end{figure}

\begin{figure}[!h]
\centering
\subfigure[\,$\pi^{-}\Sigma^{+}$]{\includegraphics[width=0.49\textwidth]{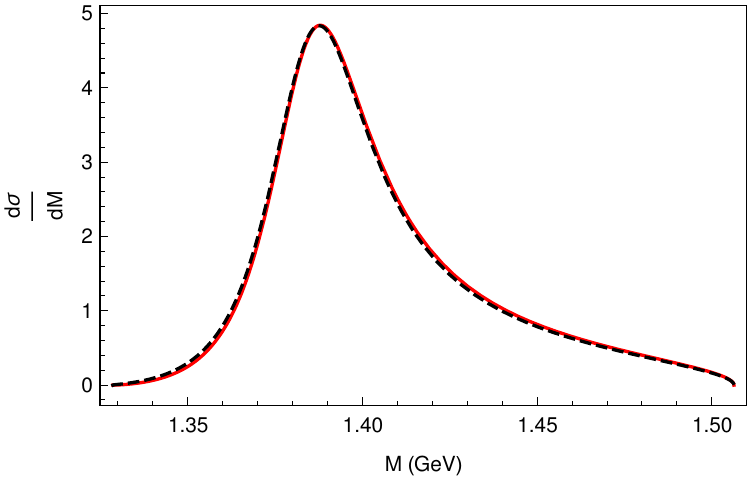}}
\subfigure[\,$\pi^{+}\Sigma^{-}$]{\includegraphics[width=0.49\textwidth]{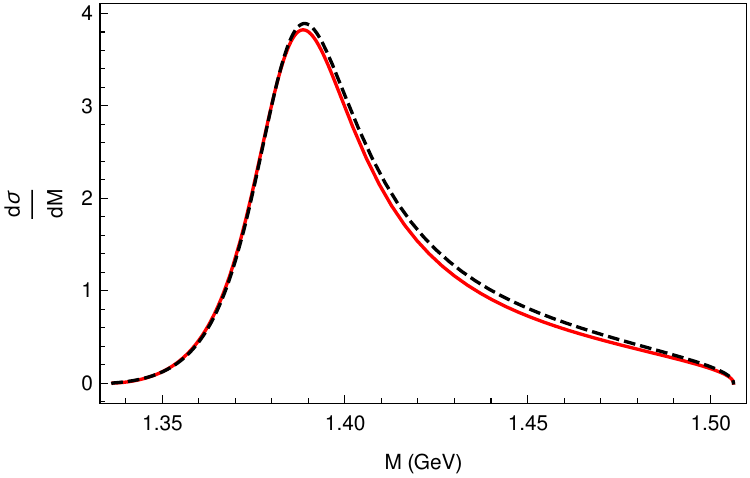}}
\caption{$\frac{d\sigma}{dM_{\pi\Sigma}}$ from $\Sigma^{\ast}$ s-channel exchange graphs for $\gamma p\rightarrow K^{+}\pi^{\mp}\Sigma^{\pm}$, $\sqrt{s}=2\,\mathrm{GeV}$, in $\mu$b/GeV. The red lines represent the ``genuine pole terms'' of Eq.~(\ref{eq:Asqr_poleapprox}), while the black dashed lines stem from the full set of graphs from App.~\ref{app:Mi_decuplet}, without partial-wave truncation. The difference between the left and right panel is mostly due to the $\Sigma^{\pm}$ mass difference.
}
\label{fig:Sigmastarexchange_ab}
\end{figure}

The corresponding squared amplitude entering the cross section then reads
\begin{eqnarray}
  |\mathcal{A}|^2_{\Sigma^{\ast}\mathrm{pole}} &=& \frac{(E_{\Sigma}^{\ast}+m_{\Sigma})|\vec{p}_{\Sigma}^{\,\ast}|^2}{36m_{D}^2}\biggl[ 8m_{D}^2(E_{N}^{\ast}+m_{N})(\mathcal{C}_{1+}^{1})^2  - 6m_{D}(s-m_{N}^2)\mathcal{C}_{1+}^{1}\tilde{\mathcal{C}}_{1+}^{5} \nonumber \\  &-& 6E_{\gamma}^{\ast}\left(2m_{D}m_{N}\mathcal{C}_{1+}^{1}\tilde{\mathcal{C}}_{1+}^{5}-3(s-m_{N}^2)\left((\tilde{\mathcal{C}}_{1+}^{5})^2 + (\tilde{\mathcal{C}}_{1+}^{6})^2\right)\right) \nonumber \\
    &+& 9(s-m_{N}^2)|\vec{q}_{K}|\sin\theta_{K}\,\mathcal{C}_{1+}^{1}\tilde{\mathcal{C}}_{1+}^{6} \nonumber \\
    &+& \frac{3(s-m_{N}^2)|\vec{q}_{K}|\sin\theta_{K}}{M_{K}^2-t_{K}}\biggl(\bigl(s-m_{N}^2+2m_{N}E_{\gamma}^{\ast}-8m_{D}(E_{N}^{\ast}+m_{N})\bigr)\mathcal{C}_{1+}^{1}\tilde{\mathcal{C}}_{1+}^{6} \nonumber \\
    & & \hspace{4cm} +\,\,2\frac{(s-m_{N}^2)}{m_{D}}|\vec{q}_{K}|\sin\theta_{K}\,\left(\mathcal{C}_{1+}^{1}\tilde{\mathcal{C}}_{1+}^{5}-3\left((\tilde{\mathcal{C}}_{1+}^{5})^2 + (\tilde{\mathcal{C}}_{1+}^{6})^2\right)\right)\biggr) \nonumber \\
    &+& \frac{4(s-m_{N}^2)^2|\vec{q}_{K}|^2\sin^2\theta_{K}}{(M_{K}^2-t_{K})^2}(E_{N}^{\ast}+m_{N})\left(\left(\mathcal{C}_{1+}^{1}-3\tilde{\mathcal{C}}_{1+}^{5}\right)^2 + 9(\tilde{\mathcal{C}}_{1+}^{6})^2\right)\biggr]\,,\label{eq:Asqr_poleapprox}
\end{eqnarray}
where we defined $\tilde{\mathcal{C}}_{1+}^{5,6}:=(m_{D}/|\vec{p}_{\Sigma}^{\,\ast}|)\mathcal{C}_{1+}^{5,6}$ for convenience. We can integrate this over the c.m. kaon angles to obtain the $\Sigma^{\ast 0}$ pole contribution to the mass spectrum:
\begin{equation}\label{eq:dsigmadM_poleapprox}
\frac{d\sigma}{dM_{\pi\Sigma}}\biggl|_{\mathrm{pole}} = \frac{|\vec{q}_{K}||\vec{p}_{\Sigma}^{\,\ast}|}{(4\pi)^4s|\vec{k}|}\int d\Omega_{K}|\mathcal{A}|^2_{\Sigma^{\ast}\mathrm{pole}}\,.
\end{equation}
Note that in this pole approximation, the $M_{\pi\Sigma}$ dependence of $|\mathcal{A}|^2_{\Sigma^{\ast}\mathrm{pole}}$ comes only from the denominator in $n_{D}(M_{\pi\Sigma})$, and from the $M_{\pi\Sigma}$ dependence of the prefactor $(E_{\Sigma}^{\ast}+m_{\Sigma})|\vec{p}_{\Sigma}^{\,\ast}|^2$ in (\ref{eq:Asqr_poleapprox}), which we retain to guarantee the correct behavior at the $\pi\Sigma$ threshold. Of course, we also retain the $M_{\pi\Sigma}$ dependence in the factor in front of the integral in (\ref{eq:dsigmadM_poleapprox}). To get a reasonable result, we include the width of the $\Sigma^{\ast 0}$ ``by hand'' in our tree-level calculation, and replace
\begin{equation}\label{pole_replacement}
\frac{1}{(m_{D}^2-M_{\pi\Sigma}^2)^2} \quad\rightarrow\quad \frac{1}{\left|m_{\Sigma^{\ast 0}}^2-iM_{\pi\Sigma}\Gamma_{\Sigma^{\ast 0}}-M_{\pi\Sigma}^2\right|^2}\,,
\end{equation}
as well as $m_{D}\rightarrow m_{\Sigma^{\ast 0}}$ everywhere in the numerator. This yields the red curve in Fig.~\ref{fig:Sigmastarexchange_poleterms}(a). \\
\quad \\
In Fig.~\ref{fig:Sigmastarexchange_poleterms}(b), we also show the dependence on the kaon angle, i.e. the differential cross section
\begin{equation}\label{eq:dsigmadz_poleapprox}
\frac{d\sigma}{dz_{K}}\biggl|_{\mathrm{pole}} = 2\pi\frac{|\vec{q}_{K}||\vec{p}_{\Sigma}^{\,\ast}|}{(4\pi)^4s|\vec{k}|}\int_{m_{\Sigma}+M_{\pi}}^{\sqrt{s}-M_{K}}dM_{\pi\Sigma}|\mathcal{A}|^2_{\Sigma^{\ast}\mathrm{pole}}\,.
\end{equation}
Next, we have to check that the approximation discussed above makes sense. To this end, we calculate the mass spectrum directly from the full invariant amplitudes in App.~\ref{app:Mi_decuplet} (i.e., from the graphs in Fig.~\ref{fig:Sigstar0exchange}), {\em without\,} performing any partial wave truncation (but using the replacement (\ref{pole_replacement}) in the $\mathcal{M}_{i}^{(a)}$). The results are shown in Fig.~\ref{fig:Sigmastarexchange_ab}. One sees that the background corrections, in particular the contributions from the $\mathcal{M}_{i}^{(b)}$ parts, are tiny in the vicinity of the resonance. If we plug in only the ``genuine pole part'' for the $\mathcal{M}_{i}$ in the full cross section formula without partial wave truncation (see Eq.~(9) in \cite{Bruns:2020lyb}, or any good textbook on high-energy physics), we get exact numerical equality to the result derived from Eqs.~(\ref{eq:Asqr_poleapprox})-(\ref{pole_replacement}) above.
%


\clearpage

\section{Summary and outlook}
\label{sec:summary}

In this note, we have provided the extension of the formalism presented in \cite{Bruns:2020lyb} to include the meson-baryon final-state interaction (FSI) with total angular momentum $J=3/2$ and orbital angular momentum $\ell=1$. As before, we have assumed that it is a reasonable approximation to exclude the $K^{+}$ from the FSI, at least for the analysis of the $\pi\Sigma$ lineshapes in the invariant-mass region where the $S=-1$ s-wave resonances dominate. The meson-baryon subsystem $MB$ produced in the intermediate reaction $\gamma p\,\rightarrow\,K^{+}MB$ undergoes FSI, $MB\,\rightarrow\,\pi\Sigma$, and this resonant FSI is mainly responsible for the prominent structures in the observed lineshapes \cite{Bruns:2022sio,Cieply:2023saa,Cieply:2022cmu}. The meson-baryon scattering amplitude describing this FSI can be decomposed into partial waves $f_{\ell\pm}(M_{\pi\Sigma})$ as in Eq.~(\ref{eq:TpwDecomp}), and we have established combinations of invariant amplitudes for the photoproduction process which satisfy partial-wave unitarity constraints of the form of Eq.~(\ref{eq:uniA}) for $J(MB)\leq 3/2$, facilitating the construction of unitary models for the photoproduction amplitude (like e.g. Eq.~(2.10) in  \cite{Bruns:2022sio}). We have found ten additional amplitude combinations for the production of $\pi\Sigma$ in a $J=3/2,\,\ell=1$ state. The explicit form of the additional terms in the cross section formula due to those ten amplitude combinations $\mathcal{C}^{j}_{1+}$  can be found in App.~\ref{app:Asqr1plus}. As a check of our projection method, we have calculated a subset of graphs with a $\Sigma^{\ast 0}$ exchange, and confirmed that only the $\mathcal{C}^{j}_{1+}$ combinations feature the $\Sigma^{\ast 0}$ pole in the $M_{\pi\Sigma}$ complex energy plane. We also found that the 'background' produced by the mentioned subset of graphs is very small in the energy region of our interest, so that the simplified expression for the cross section given by Eqs.~(\ref{eq:C1p_poleapprox})-(\ref{pole_replacement}) is a very good approximation for the contribution due to those graphs. \\
As mentioned in the introduction, the formalism used in \cite{Bruns:2020lyb,Bruns:2022sio,Cieply:2023saa} should still be improved also in other respects. In particular, the leading-order photoproduction kernel should be supplemented by terms from the next-to-leading order chiral Lagrangian and/or vector-meson exchange graphs (as done in \cite{Nakamura:2013boa}), explicit decuplet fields or other baryon resonances, or perhaps even triangle mechanisms (see \cite{Wang:2016dtb}). While those modifications will most likely lead to an improved description of the data from \cite{CLAS:2013rjt} also at slightly higher c.m. energies, it remains to be seen whether the more recent data at significantly higher c.m. energies from the GlueX collaboration \cite{Wickramaarachchi:2022mhi} can be treated with an ansatz as advocated in \cite{Bruns:2022sio,Cieply:2023saa,Cieply:2022cmu} and this work. For an application of our extended formalism to such data, we have to refer to our forthcoming publications.

\quad \\

\subsection*{Acknowledgement}

I thank Ale\v{s} Ciepl\'y for valuable comments on the manuscript. 

\clearpage

\begin{appendix}

\section{Energies and momenta in two different Lorentz frames}
\label{app:energies_momenta}
\def\theequation{\Alph{section}.\arabic{equation}}
\setcounter{equation}{0}

\qquad \\

\begin{center}
\begin{tabular}{|c|c|} 
  overall c.m. frame & $\ast$ frame \\
  \hline \qquad & \qquad \\
  $p_{N}^{\mu}+k^{\mu} = (\sqrt{s},0,0,0)$ & $p_{\Sigma}^{\mu}+q_{\pi}^{\mu} = (M_{\pi\Sigma},0,0,0)$ \\
  \qquad & \qquad \\
  $q_{K}^{\mu} = (E_{K},|\vec{q}_{K}|\sin\theta_{K},0,|\vec{q}_{K}|\cos\theta_{K})$ & $q_{K}^{\mu} = (E_{K}^{\ast},|\vec{q}_{K}^{\,\ast}|\sin\theta_{K}^{\ast},0,|\vec{q}_{K}^{\,\ast}|\cos\theta_{K}^{\ast})$ \\
  \qquad & \qquad \\
  $E_{K} = \frac{s+M_{K}^2-M_{\pi\Sigma}^2}{2\sqrt{s}}$ & $E_{K}^{\ast} = \frac{s-M_{K}^2-M_{\pi\Sigma}^2}{2M_{\pi\Sigma}}$\\
  \qquad & \qquad \\
  $|\vec{q}_{K}|=\sqrt{E_{K}^2-M_{K}^2}$ & $|\vec{q}_{K}^{\,\ast}| = \frac{\sqrt{s}}{M_{\pi\Sigma}}|\vec{q}_{K}|$ \\
  \qquad & \qquad \\
  $k^{\mu} = (E_{\gamma},0,0,E_{\gamma})$ & $k^{\mu} = (E_{\gamma}^{\ast},0,0,E_{\gamma}^{\ast})$ \\
  \qquad & \qquad \\
  $E_{\gamma} = \frac{s-m_{N}^2}{2\sqrt{s}} = |\vec{k}|$ & $E_{\gamma}^{\ast} = \frac{s-m_{N}^2 + t_{K}-M_{K}^2}{2M_{\pi\Sigma}}  = |\vec{k}^{\ast}|$ \\
  \qquad & \qquad \\
  $E_{N} = \frac{s+m_{N}^2}{2\sqrt{s}}$ & $E_{N}^{\ast} = \frac{M_{\pi\Sigma}^2 + m_{N}^2 - t_{K}}{2M_{\pi\Sigma}}$ \\
  \qquad & \qquad \\
  $\cos\theta_{K} = \frac{t_{K}-M_{K}^2 + 2E_{\gamma}E_{K}}{2E_{\gamma}|\vec{q}_{K}|} \equiv z_{K}$ & $\cos\theta_{K}^{\ast} = \frac{t_{K}-M_{K}^2 + 2E_{\gamma}^{\ast}E_{K}^{\ast}}{2E_{\gamma}^{\ast}|\vec{q}_{K}^{\,\ast}|}$\\
  \qquad & \qquad \\
  $\sin\theta_{K} = \sqrt{1-\cos^2\theta_{K}}$ & $\sin\theta_{K}^{\ast} = \frac{E_{\gamma}}{E_{\gamma}^{\ast}}\sin\theta_{K}$ \\
  \qquad & \qquad \\
  $E_{\pi\Sigma}:=E_{\pi}+E_{\Sigma} = \sqrt{s}-E_{K}$ & $E_{\pi\Sigma}^{\ast}:=E_{\pi}^{\ast}+E_{\Sigma}^{\ast} = M_{\pi\Sigma}$ \\
  \qquad & \qquad \\
  $E_{\Sigma} = \frac{2m_{\Sigma}^2 + m_{N}^2 -t_{\Sigma}- u_{\Sigma}}{2\sqrt{s}}$ & $E_{\Sigma}^{\ast}=\frac{M_{\pi\Sigma}^2+m_{\Sigma}^2-M_{\pi}^2}{2M_{\pi\Sigma}}$ \\
  \qquad & \qquad \\
  $|\vec{p}_{\Sigma}| = \sqrt{E_{\Sigma}^2-m_{\Sigma}^2}$ & $|\vec{p}_{\Sigma}^{\,\ast}| = \sqrt{E_{\Sigma}^{\ast\,2}-m_{\Sigma}^2}$ \\
  \qquad & \qquad \\ \hline
\end{tabular}
\end{center}
\qquad \\
The variables $\theta_{\Sigma}^{\ast},\,\phi_{\Sigma}^{\ast}$ over which we integrate are the standard angles in spherical coordinates, 
\begin{displaymath}
p_{\Sigma}^{\mu} = \left(E_{\Sigma}^{\ast},\,|\vec{p}_{\Sigma}^{\,\ast}|\sin\theta_{\Sigma}^{\ast}\cos\phi_{\Sigma}^{\ast},\,|\vec{p}_{\Sigma}^{\,\ast}|\sin\theta_{\Sigma}^{\ast}\sin\phi_{\Sigma}^{\ast},\,|\vec{p}_{\Sigma}^{\,\ast}|\cos\theta_{\Sigma}^{\ast}\right) \qquad \mathrm{in\,\,the\,} \ast \mathrm{\,frame}\,.
\end{displaymath}
\quad \\
$m_{N},\,m_{\Sigma},\,M_{\pi},\,M_{K}$ are, of course, the rest masses of the nucleon (a proton $p$ in our case), the $\Sigma$ hyperon, the pion and the 'spectator' kaon $K^{+}$, respectively. 
The Lorentz transformation connecting the two frames was given explicitly in App.~A of \cite{Bruns:2020lyb}\,.

\newpage

\section{Expressions for the $\mathcal{C}_{0+}^{j[1]}$, $\mathcal{C}_{1-}^{j[1]}$}
\label{app:old_Cs}
\def\theequation{\Alph{section}.\arabic{equation}}
\setcounter{equation}{0}

Here we reproduce the partial-wave projections previously derived in \cite{Bruns:2020lyb}, i.e. the parts of the $\mathcal{C}_{0+}^{j}$, $\mathcal{C}_{1-}^{j}$ which are only due to the angular averages $\overline{\mathcal{M}}_{i}\equiv \mathcal{M}_{i}^{00}$ of the invariant amplitudes:
\begin{eqnarray*}
  \mathcal{C}_{0+}^{1[1]} &=& \mathcal{M}_{1}' + (\sqrt{s}+m_{N})\mathcal{M}_{5}'+(\sqrt{s}-M_{\pi\Sigma})\left(\mathcal{M}_{9}'+(\sqrt{s}+m_{N})\mathcal{M}_{13}'\right)\,,\\
  \mathcal{C}_{0+}^{2[1]} &=& \mathcal{M}_{1}' - (\sqrt{s}-m_{N})\mathcal{M}_{5}' - (\sqrt{s}+M_{\pi\Sigma})\left(\mathcal{M}_{9}'-(\sqrt{s}-m_{N})\mathcal{M}_{13}'\right)\,,\\
  \mathcal{C}_{0+}^{3[1]} &=& \mathcal{M}_{1}' + (\sqrt{s}-M_{\pi\Sigma})\mathcal{M}_{9}' \\ &+& \frac{1}{2}(\sqrt{s}+m_{N})\left(\mathcal{M}_{2}'+(\sqrt{s}-m_{N})\mathcal{M}_{6}'+(\sqrt{s}-M_{\pi\Sigma})\left(\mathcal{M}_{10}'+(\sqrt{s}-m_{N})\mathcal{M}_{14}'\right)\right)\,,\\
  \mathcal{C}_{0+}^{4[1]} &=& \mathcal{M}_{1}' - (\sqrt{s}+M_{\pi\Sigma})\mathcal{M}_{9}' \\ &-& \frac{1}{2}(\sqrt{s}-m_{N})\left(\mathcal{M}_{2}'-(\sqrt{s}+m_{N})\mathcal{M}_{6}'-(\sqrt{s}+M_{\pi\Sigma})\left(\mathcal{M}_{10}'-(\sqrt{s}+m_{N})\mathcal{M}_{14}'\right)\right)\,,\\
  \mathcal{C}_{1-}^{1[1]} &=& \mathcal{M}_{1}''-(\sqrt{s}-m_{N})\mathcal{M}_{5}'' - (\sqrt{s}-M_{\pi\Sigma})\left(\mathcal{M}_{9}''-(\sqrt{s}-m_{N})\mathcal{M}_{13}''\right)\,,\\
  \mathcal{C}_{1-}^{2[1]} &=& \mathcal{M}_{1}'' + (\sqrt{s}+m_{N})\mathcal{M}_{5}''+(\sqrt{s}+M_{\pi\Sigma})\left(\mathcal{M}_{9}''+(\sqrt{s}+m_{N})\mathcal{M}_{13}''\right)\,,\\
  \mathcal{C}_{1-}^{3[1]} &=& \mathcal{M}_{1}'' - (\sqrt{s}-M_{\pi\Sigma})\mathcal{M}_{9}'' \\ &-& \frac{1}{2}(\sqrt{s}-m_{N})\left(\mathcal{M}_{2}''-(\sqrt{s}+m_{N})\mathcal{M}_{6}''-(\sqrt{s}-M_{\pi\Sigma})\left(\mathcal{M}_{10}''-(\sqrt{s}+m_{N})\mathcal{M}_{14}''\right)\right)\,,\\
  \mathcal{C}_{1-}^{4[1]} &=& \mathcal{M}_{1}'' + (\sqrt{s}+M_{\pi\Sigma})\mathcal{M}_{9}'' \\ &+& \frac{1}{2}(\sqrt{s}+m_{N})\left(\mathcal{M}_{2}''+(\sqrt{s}-m_{N})\mathcal{M}_{6}''+(\sqrt{s}+M_{\pi\Sigma})\left(\mathcal{M}_{10}''+(\sqrt{s}-m_{N})\mathcal{M}_{14}''\right)\right)\,,
\end{eqnarray*}
employing the following abbreviations:
\begin{eqnarray*}
  \mathcal{M}_{1}' &:=&  \overline{\mathcal{M}}_{1} - \frac{1}{3}(E_{\Sigma}^{\ast}-m_{\Sigma})\overline{\mathcal{M}}_{3}\,,\qquad \mathcal{M}_{1}'' :=  \overline{\mathcal{M}}_{1} + \frac{1}{3}(E_{\Sigma}^{\ast}+m_{\Sigma})\overline{\mathcal{M}}_{3}\,,\\
  \mathcal{M}_{5}' &:=&  \overline{\mathcal{M}}_{5} + \frac{1}{3}(E_{\Sigma}^{\ast}-m_{\Sigma})\overline{\mathcal{M}}_{7}\,,\qquad \mathcal{M}_{5}'' :=  \overline{\mathcal{M}}_{5} - \frac{1}{3}(E_{\Sigma}^{\ast}+m_{\Sigma})\overline{\mathcal{M}}_{7}\,,\\
  \mathcal{M}_{9}' &:=&  \overline{\mathcal{M}}_{9} + \frac{1}{3}(E_{\Sigma}^{\ast}-m_{\Sigma})\overline{\mathcal{M}}_{11}\,,\qquad \mathcal{M}_{9}'' :=  \overline{\mathcal{M}}_{9} - \frac{1}{3}(E_{\Sigma}^{\ast}+m_{\Sigma})\overline{\mathcal{M}}_{11}\,,\\
  \mathcal{M}_{13}' &:=&  \overline{\mathcal{M}}_{13} - \frac{1}{3}(E_{\Sigma}^{\ast}-m_{\Sigma})\overline{\mathcal{M}}_{15}\,,\qquad \mathcal{M}_{13}'' :=  \overline{\mathcal{M}}_{13} + \frac{1}{3}(E_{\Sigma}^{\ast}+m_{\Sigma})\overline{\mathcal{M}}_{15}\,,\\
  \mathcal{M}_{2}' &:=&  \overline{\mathcal{M}}_{2} + \frac{1}{3M_{\pi\Sigma}}(4E_{\Sigma}^{\ast}-m_{\Sigma})\overline{\mathcal{M}}_{3}\,,\qquad \mathcal{M}_{2}'' :=  \overline{\mathcal{M}}_{2} + \frac{1}{3M_{\pi\Sigma}}(4E_{\Sigma}^{\ast}+m_{\Sigma})\overline{\mathcal{M}}_{3}\,,\\
  \mathcal{M}_{6}' &:=&  \overline{\mathcal{M}}_{6} + \frac{1}{3M_{\pi\Sigma}}(4E_{\Sigma}^{\ast}-m_{\Sigma})\overline{\mathcal{M}}_{7}\,,\qquad \mathcal{M}_{6}'' :=  \overline{\mathcal{M}}_{6} + \frac{1}{3M_{\pi\Sigma}}(4E_{\Sigma}^{\ast}+m_{\Sigma})\overline{\mathcal{M}}_{7}\,,\\
  \mathcal{M}_{10}' &:=&  \overline{\mathcal{M}}_{10} + \frac{1}{3M_{\pi\Sigma}}(4E_{\Sigma}^{\ast}-m_{\Sigma})\overline{\mathcal{M}}_{11}\,,\qquad \mathcal{M}_{10}'' :=  \overline{\mathcal{M}}_{10} + \frac{1}{3M_{\pi\Sigma}}(4E_{\Sigma}^{\ast}+m_{\Sigma})\overline{\mathcal{M}}_{11}\,,\\
  \mathcal{M}_{14}' &:=&  \overline{\mathcal{M}}_{14} + \frac{1}{3M_{\pi\Sigma}}(4E_{\Sigma}^{\ast}-m_{\Sigma})\overline{\mathcal{M}}_{15}\,,\qquad \mathcal{M}_{14}'' :=  \overline{\mathcal{M}}_{14} + \frac{1}{3M_{\pi\Sigma}}(4E_{\Sigma}^{\ast}+m_{\Sigma})\overline{\mathcal{M}}_{15}\,.
\end{eqnarray*}
See Eq.~(\ref{eq:MmuDecomp}) for the definition of the $\mathcal{M}_{i}$, and Eq.~(\ref{eq:Mbardef}) for the definition of the $\overline{\mathcal{M}}_{i}$. We point out that the structure functions $\mathcal{M}_{4,8,12,16}$ were eliminated via the gauge-invariance constraints in Eq.~(\ref{eq:gaugeinv}), so their contributions are implicitly present in the above expressions for the $\mathcal{C}_{\ell\pm}^{i[1]}$.

\newpage

\section{Coefficients in $\Delta\mathcal{C}_{0+}$, $\Delta\mathcal{C}_{1-}$}
\label{app:cd_coefficients}
\def\theequation{\Alph{section}.\arabic{equation}}
\setcounter{equation}{0}

Here we give the coefficients $c_{\ell\pm}^{j}(k)$, $d_{\ell\pm}^{j}(k)$ appearing in Eq.~(\ref{eq:DeltaC}).

\begin{eqnarray*}
  c^{1}_{0+}(1) &=& \sqrt{s}+m_{N}-E_{\gamma}^{\ast}\,,\\
  c^{1}_{0+}(5) &=& -E_{\gamma}^{\ast}(\sqrt{s}+m_{N})\,,\\
  c^{1}_{0+}(9) &=& (\sqrt{s}+m_{N})(M_{\pi\Sigma}-m_{N})-E_{\gamma}^{\ast}(\sqrt{s}+M_{\pi\Sigma})\,,\\
  c^{1}_{0+}(13) &=& (\sqrt{s}+m_{N})\left(s-m_{N}^2-E_{\gamma}^{\ast}(\sqrt{s}+M_{\pi\Sigma})\right)\,,\\
  d^{1}_{0+}(1) &=& M_{\pi\Sigma}(s-m_{N}^2)(\sqrt{s}+m_{N}) \\ &-& E_{\gamma}^{\ast}(\sqrt{s}+m_{N})(s+M_{\pi\Sigma}^2-M_{K}^2+M_{\pi\Sigma}(\sqrt{s}-m_{N}))+2\sqrt{s}E_{\gamma}^{\ast\,2}M_{\pi\Sigma}\,,\\
  d^{1}_{0+}(5) &=& -E_{\gamma}^{\ast}(\sqrt{s}+m_{N})M_{\pi\Sigma}(s-m_{N}^2-2\sqrt{s}E_{\gamma}^{\ast})\,,\\
  d^{1}_{0+}(9) &=& M_{\pi\Sigma}(s-m_{N}^2)(\sqrt{s}+m_{N})(M_{\pi\Sigma}-m_{N})  - E_{\gamma}^{\ast}m_{N}(\sqrt{s}+m_{N})(s-m_{N}^2) \\  &-& E_{\gamma}^{\ast}(\sqrt{s}+m_{N})(M_{\pi\Sigma}-m_{N})(s+M_{\pi\Sigma}^2-M_{K}^2+(\sqrt{s}-m_{N})(\sqrt{s}+m_{N}+M_{\pi\Sigma})) \\ &+& 2\sqrt{s}E_{\gamma}^{\ast\,2}M_{\pi\Sigma}(\sqrt{s}+M_{\pi\Sigma})\,,\\
  d^{1}_{0+}(13) &=& M_{\pi\Sigma}(s-m_{N}^2)^2(\sqrt{s}+m_{N}) \\ &-& E_{\gamma}^{\ast}(s-m_{N}^2)(\sqrt{s}+m_{N})(s+M_{\pi\Sigma}^2-M_{K}^2+M_{\pi\Sigma}(\sqrt{s}+M_{\pi\Sigma})) \\ &+& 2\sqrt{s}E_{\gamma}^{\ast\,2}M_{\pi\Sigma}(\sqrt{s}+M_{\pi\Sigma})(\sqrt{s}+m_{N})\,,\\
  \quad &\quad & \quad \\
  c^{2}_{0+}(1) &=& -(\sqrt{s}-m_{N}+E_{\gamma}^{\ast})\,,\\
  c^{2}_{0+}(5) &=& E_{\gamma}^{\ast}(\sqrt{s}-m_{N})\,,\\
  c^{2}_{0+}(9) &=& -(\sqrt{s}-m_{N})(M_{\pi\Sigma}-m_{N}) +E_{\gamma}^{\ast}(\sqrt{s}-M_{\pi\Sigma})\,,\\
  c^{2}_{0+}(13) &=& -(\sqrt{s}-m_{N})\left(s-m_{N}^2+E_{\gamma}^{\ast}(\sqrt{s}-M_{\pi\Sigma})\right)\,,\\
  d^{2}_{0+}(1) &=& -M_{\pi\Sigma}(s-m_{N}^2)(\sqrt{s}-m_{N}) \\ &+& E_{\gamma}^{\ast}(\sqrt{s}-m_{N})(s+M_{\pi\Sigma}^2-M_{K}^2-M_{\pi\Sigma}(\sqrt{s}+m_{N}))-2\sqrt{s}E_{\gamma}^{\ast\,2}M_{\pi\Sigma}\,,\\
  d^{2}_{0+}(5) &=& E_{\gamma}^{\ast}(\sqrt{s}-m_{N})M_{\pi\Sigma}(s-m_{N}^2+2\sqrt{s}E_{\gamma}^{\ast})\,,\\
  d^{2}_{0+}(9) &=& -M_{\pi\Sigma}(s-m_{N}^2)(\sqrt{s}-m_{N})(M_{\pi\Sigma}-m_{N})  + E_{\gamma}^{\ast}m_{N}(\sqrt{s}-m_{N})(s-m_{N}^2) \\  &+& E_{\gamma}^{\ast}(\sqrt{s}-m_{N})(M_{\pi\Sigma}-m_{N})(s+M_{\pi\Sigma}^2-M_{K}^2+(\sqrt{s}+m_{N})(\sqrt{s}-m_{N}-M_{\pi\Sigma})) \\ &+& 2\sqrt{s}E_{\gamma}^{\ast\,2}M_{\pi\Sigma}(\sqrt{s}-M_{\pi\Sigma})\,,\\
  d^{2}_{0+}(13) &=& -M_{\pi\Sigma}(s-m_{N}^2)^2(\sqrt{s}-m_{N}) \\ &+& E_{\gamma}^{\ast}(s-m_{N}^2)(\sqrt{s}-m_{N})(s+M_{\pi\Sigma}^2-M_{K}^2-M_{\pi\Sigma}(\sqrt{s}-M_{\pi\Sigma})) \\ &-& 2\sqrt{s}E_{\gamma}^{\ast\,2}M_{\pi\Sigma}(\sqrt{s}-M_{\pi\Sigma})(\sqrt{s}-m_{N})\,,\\
  \quad &\quad & \quad \\
  c^{3}_{0+}(1) &=& -E_{\gamma}^{\ast}\,,\\
  c^{3}_{0+}(2) &=& \frac{1}{2}(\sqrt{s}+m_{N})(\sqrt{s}-m_{N}-E_{\gamma}^{\ast})\,,\\
  c^{3}_{0+}(5) &=& 0\,,\\
  c^{3}_{0+}(6) &=& -\frac{1}{2}E_{\gamma}^{\ast}(s-m_{N}^2)\,,\\
  c^{3}_{0+}(9) &=& s-m_{N}^2-E_{\gamma}^{\ast}(\sqrt{s}+M_{\pi\Sigma})\,,\\
  c^{3}_{0+}(10) &=& \frac{1}{2}(s-m_{N}^2)(M_{\pi\Sigma}+m_{N}) -\frac{1}{2}E_{\gamma}^{\ast}(\sqrt{s}+m_{N})(\sqrt{s}+M_{\pi\Sigma})\,,\\
  c^{3}_{0+}(13) &=& 0\,,\\
  c^{3}_{0+}(14) &=& \frac{1}{2}(s-m_{N}^2)(s-m_{N}^2-E_{\gamma}^{\ast}(\sqrt{s}+M_{\pi\Sigma}))\,,\\
  d^{3}_{0+}(1) &=& -E_{\gamma}^{\ast}M_{\pi\Sigma}(s-m_{N}^2-2\sqrt{s}E_{\gamma}^{\ast})\,,\\
  d^{3}_{0+}(2) &=& \frac{1}{2}M_{\pi\Sigma}(s-m_{N}^2)^2  - \frac{1}{2}E_{\gamma}^{\ast}(s-m_{N}^2)(s+M_{\pi\Sigma}^2-M_{K}^2+M_{\pi\Sigma}(\sqrt{s}+m_{N})) \\ &+& \sqrt{s}E_{\gamma}^{\ast\,2}M_{\pi\Sigma}(\sqrt{s}+m_{N})\,,\\
  d^{3}_{0+}(5) &=& 0\,,\\
  d^{3}_{0+}(6) &=& -\frac{1}{2}E_{\gamma}^{\ast}M_{\pi\Sigma}(s-m_{N}^2)(s-m_{N}^2-2\sqrt{s}E_{\gamma}^{\ast})\,,\\
  d^{3}_{0+}(9) &=& M_{\pi\Sigma}(s-m_{N}^2)^2 - E_{\gamma}^{\ast}(s-m_{N}^2)\bigl(s+M_{\pi\Sigma}^2-M_{K}^2+M_{\pi\Sigma}(\sqrt{s}+M_{\pi\Sigma})\bigr) \\ &+& 2\sqrt{s}E_{\gamma}^{\ast\,2}M_{\pi\Sigma}(\sqrt{s}+M_{\pi\Sigma})\,,\\
  d^{3}_{0+}(10) &=& \frac{1}{2}M_{\pi\Sigma}(M_{\pi\Sigma}+m_{N})(s-m_{N}^2)^2 \\ &-& \frac{1}{2}E_{\gamma}^{\ast}(s-m_{N}^2)\bigl((M_{\pi\Sigma}+m_{N})(s+M_{\pi\Sigma}^2-M_{K}^2)+M_{\pi\Sigma}(\sqrt{s}+M_{\pi\Sigma})(\sqrt{s}+m_{N})\bigr) \\ &+& \sqrt{s}E_{\gamma}^{\ast\,2}M_{\pi\Sigma}(\sqrt{s}+M_{\pi\Sigma})(\sqrt{s}+m_{N})\,,\\
  d^{3}_{0+}(13) &=& 0\,,\\
  d^{3}_{0+}(14) &=& \frac{1}{2}M_{\pi\Sigma}(s-m_{N}^2)^3   - \frac{1}{2}E_{\gamma}^{\ast}(s-m_{N}^2)^2\bigl(s+M_{\pi\Sigma}^2-M_{K}^2+M_{\pi\Sigma}(\sqrt{s}+M_{\pi\Sigma})\bigr) \\ &+& \sqrt{s}E_{\gamma}^{\ast\,2}M_{\pi\Sigma}(\sqrt{s}+M_{\pi\Sigma})(s-m_{N}^2)\,,\\
  \quad &\quad & \quad \\
  c^{4}_{0+}(1) &=& -E_{\gamma}^{\ast}\,,\\
  c^{4}_{0+}(2) &=& \frac{1}{2}(\sqrt{s}-m_{N})(\sqrt{s}+m_{N}+E_{\gamma}^{\ast})\,,\\
  c^{4}_{0+}(5) &=& 0\,,\\
  c^{4}_{0+}(6) &=& -\frac{1}{2}E_{\gamma}^{\ast}(s-m_{N}^2)\,,\\
  c^{4}_{0+}(9) &=& s-m_{N}^2+E_{\gamma}^{\ast}(\sqrt{s}-M_{\pi\Sigma})\,,\\
  c^{4}_{0+}(10) &=& \frac{1}{2}(s-m_{N}^2)(M_{\pi\Sigma}+m_{N}) -\frac{1}{2}E_{\gamma}^{\ast}(\sqrt{s}-m_{N})(\sqrt{s}-M_{\pi\Sigma})\,,\\
  c^{4}_{0+}(13) &=& 0\,,\\
  c^{4}_{0+}(14) &=& \frac{1}{2}(s-m_{N}^2)(s-m_{N}^2+E_{\gamma}^{\ast}(\sqrt{s}-M_{\pi\Sigma}))\,,\\
  d^{4}_{0+}(1) &=& -E_{\gamma}^{\ast}M_{\pi\Sigma}(s-m_{N}^2+2\sqrt{s}E_{\gamma}^{\ast})\,,\\
  d^{4}_{0+}(2) &=& \frac{1}{2}M_{\pi\Sigma}(s-m_{N}^2)^2  - \frac{1}{2}E_{\gamma}^{\ast}(s-m_{N}^2)(s+M_{\pi\Sigma}^2-M_{K}^2-M_{\pi\Sigma}(\sqrt{s}-m_{N})) \\ &+& \sqrt{s}E_{\gamma}^{\ast\,2}M_{\pi\Sigma}(\sqrt{s}-m_{N})\,,\\
  d^{4}_{0+}(5) &=& 0\,,\\
  d^{4}_{0+}(6) &=& -\frac{1}{2}E_{\gamma}^{\ast}M_{\pi\Sigma}(s-m_{N}^2)(s-m_{N}^2+2\sqrt{s}E_{\gamma}^{\ast})\,,\\
  d^{4}_{0+}(9) &=& M_{\pi\Sigma}(s-m_{N}^2)^2 - E_{\gamma}^{\ast}(s-m_{N}^2)\bigl(s+M_{\pi\Sigma}^2-M_{K}^2-M_{\pi\Sigma}(\sqrt{s}-M_{\pi\Sigma})\bigr) \\ &+& 2\sqrt{s}E_{\gamma}^{\ast\,2}M_{\pi\Sigma}(\sqrt{s}-M_{\pi\Sigma})\,,\\
  d^{4}_{0+}(10) &=& \frac{1}{2}M_{\pi\Sigma}(M_{\pi\Sigma}+m_{N})(s-m_{N}^2)^2 \\ &-& \frac{1}{2}E_{\gamma}^{\ast}(s-m_{N}^2)\bigl((M_{\pi\Sigma}+m_{N})(s+M_{\pi\Sigma}^2-M_{K}^2)+M_{\pi\Sigma}(\sqrt{s}-M_{\pi\Sigma})(\sqrt{s}-m_{N})\bigr) \\ &-& \sqrt{s}E_{\gamma}^{\ast\,2}M_{\pi\Sigma}(\sqrt{s}-M_{\pi\Sigma})(\sqrt{s}-m_{N})\,,\\
  d^{4}_{0+}(13) &=& 0\,,\\
  d^{4}_{0+}(14) &=& \frac{1}{2}M_{\pi\Sigma}(s-m_{N}^2)^3   - \frac{1}{2}E_{\gamma}^{\ast}(s-m_{N}^2)^2\bigl(s+M_{\pi\Sigma}^2-M_{K}^2-M_{\pi\Sigma}(\sqrt{s}-M_{\pi\Sigma})\bigr) \\ &+& \sqrt{s}E_{\gamma}^{\ast\,2}M_{\pi\Sigma}(\sqrt{s}-M_{\pi\Sigma})(s-m_{N}^2)\,,
\end{eqnarray*}
\begin{eqnarray*}
  c^{1}_{1-}(1) &=& \sqrt{s}-m_{N}-E_{\gamma}^{\ast}\,,\\
  c^{1}_{1-}(5) &=& E_{\gamma}^{\ast}(\sqrt{s}-m_{N})\,,\\
  c^{1}_{1-}(9) &=& -(\sqrt{s}-m_{N})(M_{\pi\Sigma}+m_{N})+E_{\gamma}^{\ast}(\sqrt{s}+M_{\pi\Sigma})\,,\\
  c^{1}_{1-}(13) &=& (\sqrt{s}-m_{N})\left(s-m_{N}^2-E_{\gamma}^{\ast}(\sqrt{s}+M_{\pi\Sigma})\right)\,,\\
  d^{1}_{1-}(1) &=& M_{\pi\Sigma}(s-m_{N}^2)(\sqrt{s}-m_{N}) \\ &-& E_{\gamma}^{\ast}(\sqrt{s}-m_{N})(s+M_{\pi\Sigma}^2-M_{K}^2+M_{\pi\Sigma}(\sqrt{s}+m_{N}))+2\sqrt{s}E_{\gamma}^{\ast\,2}M_{\pi\Sigma}\,,\\
  d^{1}_{1-}(5) &=& E_{\gamma}^{\ast}(\sqrt{s}-m_{N})M_{\pi\Sigma}(s-m_{N}^2-2\sqrt{s}E_{\gamma}^{\ast})\,,\\
  d^{1}_{1-}(9) &=& -M_{\pi\Sigma}(s-m_{N}^2)(\sqrt{s}-m_{N})(M_{\pi\Sigma}+m_{N})  - E_{\gamma}^{\ast}m_{N}(\sqrt{s}-m_{N})(s-m_{N}^2) \\  &+& E_{\gamma}^{\ast}(\sqrt{s}-m_{N})(M_{\pi\Sigma}+m_{N})(s+M_{\pi\Sigma}^2-M_{K}^2+(\sqrt{s}+m_{N})(\sqrt{s}-m_{N}+M_{\pi\Sigma})) \\ &-& 2\sqrt{s}E_{\gamma}^{\ast\,2}M_{\pi\Sigma}(\sqrt{s}+M_{\pi\Sigma})\,,\\
  d^{1}_{1-}(13) &=& M_{\pi\Sigma}(s-m_{N}^2)^2(\sqrt{s}-m_{N}) \\ &-& E_{\gamma}^{\ast}(s-m_{N}^2)(\sqrt{s}-m_{N})(s+M_{\pi\Sigma}^2-M_{K}^2+M_{\pi\Sigma}(\sqrt{s}+M_{\pi\Sigma})) \\ &+& 2\sqrt{s}E_{\gamma}^{\ast\,2}M_{\pi\Sigma}(\sqrt{s}+M_{\pi\Sigma})(\sqrt{s}-m_{N})\,,\\
  \quad &\quad & \quad \\
  c^{2}_{1-}(1) &=& -(\sqrt{s}+m_{N}+E_{\gamma}^{\ast})\,,\\
  c^{2}_{1-}(5) &=& -E_{\gamma}^{\ast}(\sqrt{s}+m_{N})\,,\\
  c^{2}_{1-}(9) &=& (\sqrt{s}+m_{N})(M_{\pi\Sigma}+m_{N}) -E_{\gamma}^{\ast}(\sqrt{s}-M_{\pi\Sigma})\,,\\
  c^{2}_{1-}(13) &=& -(\sqrt{s}+m_{N})\left(s-m_{N}^2+E_{\gamma}^{\ast}(\sqrt{s}-M_{\pi\Sigma})\right)\,,\\
  d^{2}_{1-}(1) &=& -M_{\pi\Sigma}(s-m_{N}^2)(\sqrt{s}+m_{N}) \\ &+& E_{\gamma}^{\ast}(\sqrt{s}+m_{N})(s+M_{\pi\Sigma}^2-M_{K}^2-M_{\pi\Sigma}(\sqrt{s}-m_{N}))-2\sqrt{s}E_{\gamma}^{\ast\,2}M_{\pi\Sigma}\,,\\
  d^{2}_{1-}(5) &=& -E_{\gamma}^{\ast}(\sqrt{s}+m_{N})M_{\pi\Sigma}(s-m_{N}^2+2\sqrt{s}E_{\gamma}^{\ast})\,,\\
  d^{2}_{1-}(9) &=& M_{\pi\Sigma}(s-m_{N}^2)(\sqrt{s}+m_{N})(M_{\pi\Sigma}+m_{N})  + E_{\gamma}^{\ast}m_{N}(\sqrt{s}+m_{N})(s-m_{N}^2) \\  &-& E_{\gamma}^{\ast}(\sqrt{s}+m_{N})(M_{\pi\Sigma}+m_{N})(s+M_{\pi\Sigma}^2-M_{K}^2+(\sqrt{s}-m_{N})(\sqrt{s}+m_{N}-M_{\pi\Sigma})) \\ &-& 2\sqrt{s}E_{\gamma}^{\ast\,2}M_{\pi\Sigma}(\sqrt{s}-M_{\pi\Sigma})\,,\\
  d^{2}_{1-}(13) &=& -M_{\pi\Sigma}(s-m_{N}^2)^2(\sqrt{s}+m_{N}) \\ &+& E_{\gamma}^{\ast}(s-m_{N}^2)(\sqrt{s}+m_{N})(s+M_{\pi\Sigma}^2-M_{K}^2-M_{\pi\Sigma}(\sqrt{s}-M_{\pi\Sigma})) \\ &-& 2\sqrt{s}E_{\gamma}^{\ast\,2}M_{\pi\Sigma}(\sqrt{s}-M_{\pi\Sigma})(\sqrt{s}+m_{N})\,,\\
  \quad &\quad & \quad \\
  c^{3}_{1-}(1) &=& -E_{\gamma}^{\ast}\,,\\
  c^{3}_{1-}(2) &=& -\frac{1}{2}(\sqrt{s}-m_{N})(\sqrt{s}+m_{N}-E_{\gamma}^{\ast})\,,\\
  c^{3}_{1-}(5) &=& 0\,,\\
  c^{3}_{1-}(6) &=& -\frac{1}{2}E_{\gamma}^{\ast}(s-m_{N}^2)\,,\\
  c^{3}_{1-}(9) &=& -\bigl(s-m_{N}^2-E_{\gamma}^{\ast}(\sqrt{s}+M_{\pi\Sigma})\bigr)\,,\\
  c^{3}_{1-}(10) &=& \frac{1}{2}(s-m_{N}^2)(M_{\pi\Sigma}-m_{N}) -\frac{1}{2}E_{\gamma}^{\ast}(\sqrt{s}-m_{N})(\sqrt{s}+M_{\pi\Sigma})\,,\\
  c^{3}_{1-}(13) &=& 0\,,\\
  c^{3}_{1-}(14) &=& -\frac{1}{2}(s-m_{N}^2)(s-m_{N}^2-E_{\gamma}^{\ast}(\sqrt{s}+M_{\pi\Sigma}))\,,\\
  d^{3}_{1-}(1) &=& -E_{\gamma}^{\ast}M_{\pi\Sigma}(s-m_{N}^2-2\sqrt{s}E_{\gamma}^{\ast})\,,\\
  d^{3}_{1-}(2) &=& -\frac{1}{2}M_{\pi\Sigma}(s-m_{N}^2)^2  + \frac{1}{2}E_{\gamma}^{\ast}(s-m_{N}^2)(s+M_{\pi\Sigma}^2-M_{K}^2+M_{\pi\Sigma}(\sqrt{s}-m_{N})) \\ &-& \sqrt{s}E_{\gamma}^{\ast\,2}M_{\pi\Sigma}(\sqrt{s}-m_{N})\,,\\
  d^{3}_{1-}(5) &=& 0\,,\\
  d^{3}_{1-}(6) &=& -\frac{1}{2}E_{\gamma}^{\ast}M_{\pi\Sigma}(s-m_{N}^2)(s-m_{N}^2-2\sqrt{s}E_{\gamma}^{\ast})\,,\\
  d^{3}_{1-}(9) &=& -M_{\pi\Sigma}(s-m_{N}^2)^2 + E_{\gamma}^{\ast}(s-m_{N}^2)\bigl(s+M_{\pi\Sigma}^2-M_{K}^2+M_{\pi\Sigma}(\sqrt{s}+M_{\pi\Sigma})\bigr) \\ &-& 2\sqrt{s}E_{\gamma}^{\ast\,2}M_{\pi\Sigma}(\sqrt{s}+M_{\pi\Sigma})\,,\\
  d^{3}_{1-}(10) &=& \frac{1}{2}M_{\pi\Sigma}(M_{\pi\Sigma}-m_{N})(s-m_{N}^2)^2 \\ &-& \frac{1}{2}E_{\gamma}^{\ast}(s-m_{N}^2)\bigl((M_{\pi\Sigma}-m_{N})(s+M_{\pi\Sigma}^2-M_{K}^2)+M_{\pi\Sigma}(\sqrt{s}+M_{\pi\Sigma})(\sqrt{s}-m_{N})\bigr) \\ &+& \sqrt{s}E_{\gamma}^{\ast\,2}M_{\pi\Sigma}(\sqrt{s}+M_{\pi\Sigma})(\sqrt{s}-m_{N})\,,\\
  d^{3}_{1-}(13) &=& 0\,,\\
  d^{3}_{1-}(14) &=& -\frac{1}{2}M_{\pi\Sigma}(s-m_{N}^2)^3  + \frac{1}{2}E_{\gamma}^{\ast}(s-m_{N}^2)^2\bigl(s+M_{\pi\Sigma}^2-M_{K}^2+M_{\pi\Sigma}(\sqrt{s}+M_{\pi\Sigma})\bigr) \\ &-& \sqrt{s}E_{\gamma}^{\ast\,2}M_{\pi\Sigma}(\sqrt{s}+M_{\pi\Sigma})(s-m_{N}^2)\,,\\
  \quad &\quad & \quad \\
  c^{4}_{1-}(1) &=& -E_{\gamma}^{\ast}\,,\\
  c^{4}_{1-}(2) &=& -\frac{1}{2}(\sqrt{s}+m_{N})(\sqrt{s}-m_{N}+E_{\gamma}^{\ast})\,,\\
  c^{4}_{1-}(5) &=& 0\,,\\
  c^{4}_{1-}(6) &=& -\frac{1}{2}E_{\gamma}^{\ast}(s-m_{N}^2)\,,\\
  c^{4}_{1-}(9) &=& -\bigl(s-m_{N}^2+E_{\gamma}^{\ast}(\sqrt{s}-M_{\pi\Sigma})\bigr)\,,\\
  c^{4}_{1-}(10) &=& \frac{1}{2}(s-m_{N}^2)(M_{\pi\Sigma}-m_{N}) -\frac{1}{2}E_{\gamma}^{\ast}(\sqrt{s}+m_{N})(\sqrt{s}-M_{\pi\Sigma})\,,\\
  c^{4}_{1-}(13) &=& 0\,,\\
  c^{4}_{1-}(14) &=& -\frac{1}{2}(s-m_{N}^2)(s-m_{N}^2+E_{\gamma}^{\ast}(\sqrt{s}-M_{\pi\Sigma}))\,,\\
  d^{4}_{1-}(1) &=& -E_{\gamma}^{\ast}M_{\pi\Sigma}(s-m_{N}^2+2\sqrt{s}E_{\gamma}^{\ast})\,,\\
  d^{4}_{1-}(2) &=& -\frac{1}{2}M_{\pi\Sigma}(s-m_{N}^2)^2  + \frac{1}{2}E_{\gamma}^{\ast}(s-m_{N}^2)(s+M_{\pi\Sigma}^2-M_{K}^2-M_{\pi\Sigma}(\sqrt{s}+m_{N})) \\ &-& \sqrt{s}E_{\gamma}^{\ast\,2}M_{\pi\Sigma}(\sqrt{s}+m_{N})\,,\\
  d^{4}_{1-}(5) &=& 0\,,\\
  d^{4}_{1-}(6) &=& -\frac{1}{2}E_{\gamma}^{\ast}M_{\pi\Sigma}(s-m_{N}^2)(s-m_{N}^2+2\sqrt{s}E_{\gamma}^{\ast})\,,\\
  d^{4}_{1-}(9) &=& -M_{\pi\Sigma}(s-m_{N}^2)^2 + E_{\gamma}^{\ast}(s-m_{N}^2)\bigl(s+M_{\pi\Sigma}^2-M_{K}^2-M_{\pi\Sigma}(\sqrt{s}-M_{\pi\Sigma})\bigr) \\ &-& 2\sqrt{s}E_{\gamma}^{\ast\,2}M_{\pi\Sigma}(\sqrt{s}-M_{\pi\Sigma})\,,\\
  d^{4}_{1-}(10) &=& \frac{1}{2}M_{\pi\Sigma}(M_{\pi\Sigma}-m_{N})(s-m_{N}^2)^2 \\ &-& \frac{1}{2}E_{\gamma}^{\ast}(s-m_{N}^2)\bigl((M_{\pi\Sigma}-m_{N})(s+M_{\pi\Sigma}^2-M_{K}^2)+M_{\pi\Sigma}(\sqrt{s}-M_{\pi\Sigma})(\sqrt{s}+m_{N})\bigr) \\ &-& \sqrt{s}E_{\gamma}^{\ast\,2}M_{\pi\Sigma}(\sqrt{s}-M_{\pi\Sigma})(\sqrt{s}+m_{N})\,,\\
  d^{4}_{1-}(13) &=& 0\,,\\
  d^{4}_{1-}(14) &=& -\frac{1}{2}M_{\pi\Sigma}(s-m_{N}^2)^3  + \frac{1}{2}E_{\gamma}^{\ast}(s-m_{N}^2)^2\bigl(s+M_{\pi\Sigma}^2-M_{K}^2-M_{\pi\Sigma}(\sqrt{s}-M_{\pi\Sigma})\bigr) \\ &-& \sqrt{s}E_{\gamma}^{\ast\,2}M_{\pi\Sigma}(\sqrt{s}-M_{\pi\Sigma})(s-m_{N}^2)\,.
\end{eqnarray*}
\quad \\
All $c_{\ell\pm}^{j}(k)$, $d_{\ell\pm}^{j}(k)$ of Eq.~(\ref{eq:DeltaC}) not listed above are to be set to zero. 

\newpage

\section{Explicit formula for $|\mathcal{A}|^2_{1+}$}
\label{app:Asqr1plus}
\def\theequation{\Alph{section}.\arabic{equation}}
\setcounter{equation}{0}

The explicit expression for the partial-wave cross section for $J(\pi\Sigma)=\frac{3}{2}$, $\ell(\pi\Sigma)=1$ is rather unwieldy in general. As we have seen in Sec.~\ref{sec:pwp}, there are $10$ different $MB(1+)$ amplitudes for two-meson photoproduction, in contrast to one ($f_{1+}$) for meson-baryon scattering, or two (multipole amplitudes $E_{1+}$, $M_{1+}$) for single-meson photoproduction, which lets us expect a hundred terms in the cross section formula. These ten amplitudes are related to the invariant amplitudes for the photoproduction of a spin $\frac{3}{2}$ baryon and a pseudoscalar meson, just like the four $\mathcal{A}^{i}_{0+}$ are related to the four independent invariant amplitudes for $\gamma p\,\rightarrow K^{+}\Lambda(1405)$ (compare the discussion in Sec.~4 of \cite{Bruns:2020lyb}). \\
Since $|\mathcal{A}|^2_{1+}$ in Eq.~(\ref{eq:d2csA}) is clearly a real bilinear expression in the $\mathcal{C}_{1+}^{j}$, we form a ten-component column vector $\mathcal{V}_{1+}$ with entries $\sqrt{E_{\Sigma}^{\ast}+m_{\Sigma}}\,\mathcal{C}_{1+}^{j}$, $j=1,\ldots,10$, for a fixed channel $MB=\pi\Sigma$ (so this is {\em not\,} a ``channel vector''), 
\begin{displaymath}
\mathcal{V}_{1+} = (\sqrt{E_{\Sigma}^{\ast}+m_{\Sigma}}\,\mathcal{C}_{1+}^{1},\,\sqrt{E_{\Sigma}^{\ast}+m_{\Sigma}}\,\mathcal{C}_{1+}^{2},\,\ldots,\,\sqrt{E_{\Sigma}^{\ast}+m_{\Sigma}}\,\mathcal{C}_{1+}^{10}\,)^{\top}\,,
\end{displaymath}
and write 
\begin{equation}\label{eq:vGv}
|\mathcal{A}|^2_{1+} = (\mathcal{V}_{1+})^{\dagger}\mathcal{G}_{1+}\mathcal{V}_{1+}\,.
\end{equation}
Here $\mathcal{G}_{1+}$ is a symmetric $10\times 10$ matrix with real entries $g_{ij}=g_{ji}\,$:
\begin{eqnarray*}
  g_{11} &=& \frac{|\vec{p}_{\Sigma}^{\,\ast}|^2}{9}\left(2+\frac{(s-m_{N}^2)^2}{(M_{K}^2-t_{K})^2}\frac{|\vec{q}_{K}|^2}{M_{\pi\Sigma}^2}\sin^2\theta_{K}\right)(E_{N}^{\ast} + m_{N})\,,\\
  g_{12} &=& \frac{|\vec{p}_{\Sigma}^{\,\ast}|^2}{18}\left(2+\frac{(s-m_{N}^2)^2}{(M_{K}^2-t_{K})^2}\frac{|\vec{q}_{K}|^2}{M_{\pi\Sigma}^2}\sin^2\theta_{K}\right) \left(s-m_{N}^2 + 2m_{N}E_{\gamma}^{\ast}\right)\,,\\
  g_{13} &=& \frac{|\vec{p}_{\Sigma}^{\,\ast}|}{6}\biggl[2E_{N}^{\ast}-\frac{\sqrt{s}E_{\gamma}}{E_{\gamma}^{\ast}} + (2m_{N}+M_{\pi\Sigma})\frac{s}{M_{\pi\Sigma}^2} \\ &+& \frac{s-m_{N}^2}{M_{\pi\Sigma}^2(M_{K}^2-t_{K})^2}\biggl((t_{K}-M_{K}^2)\left(2(m_{N}+M_{\pi\Sigma})(s-M_{\pi\Sigma}^2)+M_{K}^2(2m_{N}+M_{\pi\Sigma})\right) \\ &\quad & \hspace{8cm} + \,\, 2M_{K}^2(s-m_{N}^2)(m_{N}+M_{\pi\Sigma}) \biggr)\biggr]\,,\\
  g_{14} &=& \frac{|\vec{p}_{\Sigma}^{\,\ast}|}{6}\biggl[\frac{\sqrt{s}E_{\gamma}}{M_{\pi\Sigma}E_{\gamma}^{\ast}} - \frac{s-m_{N}^2}{2M_{\pi\Sigma}^2(M_{K}^2-t_{K})^2}\biggl((M_{K}^2-t_{K})^2 \\ & & \qquad\,\,+\,\, (M_{\pi\Sigma}^2+(2M_{\pi\Sigma}+m_{N})^2 -2s-M_{K}^2)(M_{K}^2-t_{K}) +2M_{K}^2(s-m_{N}^2)\biggr)\biggr]|\vec{q}_{K}|\sin\theta_{K}\,,\\
  g_{15} &=& -\frac{|\vec{p}_{\Sigma}^{\,\ast}|}{3}\frac{(s-m_{N}^2)^2}{(t_{K}-M_{K}^2)^2}\frac{|\vec{q}_{K}|^2}{M_{\pi\Sigma}}\sin^2\theta_{K}(E_{N}^{\ast} + m_{N})\,,\\
  g_{16} &=& \frac{|\vec{p}_{\Sigma}^{\,\ast}|}{3}\frac{s-m_{N}^2}{t_{K}-M_{K}^2}|\vec{q}_{K}|\sin\theta_{K}(E_{N}^{\ast} + m_{N})\,,\\
  g_{17} &=& -\frac{|\vec{p}_{\Sigma}^{\,\ast}|}{6}\biggl[s-m_{N}^2+2m_{N}E^{\ast}_{\gamma} + \frac{(s-m_{N}^2)^2}{t_{K}-M_{K}^2}\frac{|\vec{q}_{K}|^2}{M_{\pi\Sigma}^2}\sin^2\theta_{K}\biggr]\,,\\
  g_{18} &=& \frac{|\vec{p}_{\Sigma}^{\,\ast}|}{12}\frac{s-m_{N}^2}{t_{K}-M_{K}^2}\frac{|\vec{q}_{K}|}{M_{\pi\Sigma}}\sin\theta_{K}\,\left[3(t_{K}-M_{K}^2)-(s-m_{N}^2+2m_{N}E^{\ast}_{\gamma})\right]\,,\\
  g_{19} &=& -\frac{|\vec{p}_{\Sigma}^{\,\ast}|(s-m_{N}^2)^2}{12M_{\pi\Sigma}^2(M_{K}^2-t_{K})^2}\biggl[2(M_{\pi\Sigma}+m_{N})(s-m_{N}^2) + (M_{K}^2-t_{K})(M_{\pi\Sigma}-2m_{N})\biggr] |\vec{q}_{K}|^2\sin^2\theta_{K}\,\,,\\
  g_{110} &=& \frac{|\vec{p}_{\Sigma}^{\,\ast}|(s-m_{N}^2)}{12M_{\pi\Sigma}^2(M_{K}^2-t_{K})^2}\biggl[(s-m_{N}^2)^2|\vec{q}_{K}|^2\sin^2\theta_{K} \\ & & \hspace{2.6cm} -\,\,2M_{\pi\Sigma}(M_{K}^2-t_{K})((s-m_{N}^2)(M_{\pi\Sigma}+m_{N}) - m_{N}(M_{K}^2-t_{K}))\biggr] |\vec{q}_{K}|\sin\theta_{K}\,\,,\\
  g_{22} &=& \frac{|\vec{p}_{\Sigma}^{\,\ast}|^2}{9}\left(2+\frac{(s-m_{N}^2)^2}{(M_{K}^2-t_{K})^2}\frac{|\vec{q}_{K}|^2}{M_{\pi\Sigma}^2}\sin^2\theta_{K}\right) E_{\gamma}^{\ast}(s-m_{N}^2)\,,\\
  g_{23} &=&  \frac{|\vec{p}_{\Sigma}^{\,\ast}|}{3}\frac{E_{\gamma}^{\ast}}{M_{\pi\Sigma}}\biggl[m_{N}M_{\pi\Sigma} - s + \frac{M_{K}^2(s-m_{N}^2)}{M_{K}^2-t_{K}} -2\frac{(s-m_{N}^2)^2}{(M_{K}^2-t_{K})^2}|\vec{q}_{K}|^2\sin^2\theta_{K} \biggr]\,,\\
  g_{24} &=&  \frac{|\vec{p}_{\Sigma}^{\,\ast}|}{6}(3M_{\pi\Sigma}+m_{N})\frac{E_{\gamma}^{\ast}}{M_{\pi\Sigma}}\biggl[\frac{s-m_{N}^2}{t_{K}-M_{K}^2}|\vec{q}_{K}|\sin\theta_{K} \biggr]\,,\\
  g_{25} &=& \frac{|\vec{p}_{\Sigma}^{\,\ast}|}{12}\frac{(s-m_{N}^2)^2}{(t_{K}-M_{K}^2)^2}\frac{|\vec{q}_{K}|^2}{M_{\pi\Sigma}}\sin^2\theta_{K}\left[M_{K}^2-t_{K}-2(s-m_{N}^2+2m_{N}E^{\ast}_{\gamma})\right]\,,\\
  g_{26} &=& \frac{|\vec{p}_{\Sigma}^{\,\ast}|}{12}\biggl[2M_{\pi\Sigma}E_{\gamma}^{\ast}\biggl(\frac{M_{K}^2(s-m_{N}^2)}{(M_{K}^2-t_{K})^2} + \frac{s-m_{N}^2+(M_{\pi\Sigma}+m_{N})^2}{t_{K}-M_{K}^2}\biggr) - M_{\pi\Sigma}^2\biggr] (s-m_{N}^2)\frac{|\vec{q}_{K}|}{M_{\pi\Sigma}^2}\sin\theta_{K}\,,\\
  g_{27} &=& -\frac{|\vec{p}_{\Sigma}^{\,\ast}|}{3}E_{\gamma}^{\ast}(s-m_{N}^2)\,,\\
  g_{28} &=& \frac{|\vec{p}_{\Sigma}^{\,\ast}|}{6}E_{\gamma}^{\ast}\frac{(s-m_{N}^2)^2}{M_{K}^2-t_{K}}\frac{|\vec{q}_{K}|}{M_{\pi\Sigma}}\sin\theta_{K}\,,\\
  g_{29} &=& -\frac{|\vec{p}_{\Sigma}^{\,\ast}|(s-m_{N}^2)^3}{3M_{\pi\Sigma}(M_{K}^2-t_{K})^2}E_{\gamma}^{\ast}|\vec{q}_{K}|^2\sin^2\theta_{K}\,,\\
  g_{210} &=& -\frac{|\vec{p}_{\Sigma}^{\,\ast}|(s-m_{N}^2)^2}{3(M_{K}^2-t_{K})}E_{\gamma}^{\ast} |\vec{q}_{K}|\sin\theta_{K}\,,\\
  g_{33} &=& \frac{1}{M_{\pi\Sigma}(M_{K}^2-t_{K})^2}\biggl((M_{K}^2-t_{K})^3 + ((M_{\pi\Sigma}-m_{N})^2-2s-M_{K}^2)(M_{K}^2-t_{K})^2 \\ &\quad& \hspace{5.5cm}  +\,\,8sE_{\gamma}E_{K}(M_{K}^2-t_{K}) - 2M_{K}^2(s-m_{N}^2)^2\biggr)\,,\\
  g_{34} &=& 0\,,\\
  g_{35} &=& \frac{M_{\pi\Sigma}+m_{N}}{M_{\pi\Sigma}}\frac{(s-m_{N}^2)^2}{(M_{K}^2-t_{K})^2}|\vec{q}_{K}|^2\sin^2\theta_{K}\,,\\
  g_{36} &=& -\frac{(s-m_{N}^2)|\vec{q}_{K}|\sin\theta_{K}}{4M_{\pi\Sigma}}\biggl[1-\frac{2s-M_{\pi\Sigma}^2-m_{N}^2+M_{K}^2}{M_{K}^2-t_{K}} +2M_{K}^2\frac{s-m_{N}^2}{(M_{K}^2-t_{K})^2}\biggr]\,,\\
  g_{37} &=& s-m_{N}^2-2m_{N}E_{\gamma}^{\ast}\,,\\
  g_{38} &=& \frac{(s-m_{N}^2)^2|\vec{q}_{K}|\sin\theta_{K}}{2M_{\pi\Sigma}(t_{K}-M_{K}^2)}\,,\\
  g_{39} &=& (s-m_{N}^2+2M_{\pi\Sigma}E_{\gamma}^{\ast})\frac{(s-m_{N}^2)^2|\vec{q}_{K}|^2\sin^2\theta_{K}}{2M_{\pi\Sigma}(M_{K}^2-t_{K})^2}\,,\\
  g_{310} &=& (s-m_{N}^2-2m_{N}E_{\gamma}^{\ast})\frac{(s-m_{N}^2)|\vec{q}_{K}|\sin\theta_{K}}{4(M_{K}^2-t_{K})}\,,\\
  g_{44} &=& g_{33}\,,\\
  g_{45} &=& \frac{(s-m_{N}^2)|\vec{q}_{K}|\sin\theta_{K}}{4M_{\pi\Sigma}}\biggl[1-\frac{2s-M_{\pi\Sigma}^2-m_{N}^2+M_{K}^2}{M_{K}^2-t_{K}} +2M_{K}^2\frac{s-m_{N}^2}{(M_{K}^2-t_{K})^2}\biggr]\,,\\
  g_{46} &=& \frac{M_{\pi\Sigma}+m_{N}}{M_{\pi\Sigma}}\frac{(s-m_{N}^2)^2}{(M_{K}^2-t_{K})^2}|\vec{q}_{K}|^2\sin^2\theta_{K}\,,\\
  g_{47} &=& \frac{(s-m_{N}^2)^2|\vec{q}_{K}|\sin\theta_{K}}{2M_{\pi\Sigma}(M_{K}^2-t_{K})}\,,\\
  g_{48} &=& s-m_{N}^2-2m_{N}E_{\gamma}^{\ast}\,,\\
  g_{49} &=& (s-m_{N}^2-2m_{N}E_{\gamma}^{\ast})\frac{(s-m_{N}^2)|\vec{q}_{K}|\sin\theta_{K}}{4(t_{K}-M_{K}^2)}\,,\\
  g_{410} &=& (s-m_{N}^2+2M_{\pi\Sigma}E_{\gamma}^{\ast})\frac{(s-m_{N}^2)^2|\vec{q}_{K}|^2\sin^2\theta_{K}}{2M_{\pi\Sigma}(M_{K}^2-t_{K})^2}\,,\\
  g_{55} &=& \frac{(s-m_{N}^2)^2}{(M_{K}^2-t_{K})^2}|\vec{q}_{K}|^2\sin^2\theta_{K}(E_{N}^{\ast} + m_{N})\,,\\
  g_{56} &=& 0\,,\\
  g_{57} &=& \frac{(s-m_{N}^2)^2}{2M_{\pi\Sigma}(t_{K}-M_{K}^2)}|\vec{q}_{K}|^2\sin^2\theta_{K}\,,\\
  g_{58} &=& \frac{s-m_{N}^2}{4(t_{K}-M_{K}^2)}|\vec{q}_{K}|\sin\theta_{K}\left(s-m_{N}^2 + 2m_{N}E_{\gamma}^{\ast}\right)\,,\\
  g_{59} &=& (s-m_{N}^2+2m_{N}E_{\gamma}^{\ast})\frac{(s-m_{N}^2)^2|\vec{q}_{K}|^2\sin^2\theta_{K}}{2(M_{K}^2-t_{K})^2}\,,\\
  g_{510} &=& -\frac{(s-m_{N}^2)^3|\vec{q}_{K}|^3\sin^3\theta_{K}}{4M_{\pi\Sigma}(M_{K}^2-t_{K})^2}\,,\\
  g_{66} &=& g_{55}\,,\\
  g_{67} &=& -\frac{s-m_{N}^2}{4(t_{K}-M_{K}^2)}|\vec{q}_{K}|\sin\theta_{K}\left(s-m_{N}^2 + 2m_{N}E_{\gamma}^{\ast}\right)\,,\\
  g_{68} &=& \frac{(s-m_{N}^2)^2}{2M_{\pi\Sigma}(t_{K}-M_{K}^2)}|\vec{q}_{K}|^2\sin^2\theta_{K}\,,\\
  g_{69} &=& \frac{(s-m_{N}^2)^3|\vec{q}_{K}|^3\sin^3\theta_{K}}{4M_{\pi\Sigma}(M_{K}^2-t_{K})^2}\,,\\
  g_{610} &=& (s-m_{N}^2+2m_{N}E_{\gamma}^{\ast})\frac{(s-m_{N}^2)^2|\vec{q}_{K}|^2\sin^2\theta_{K}}{2(M_{K}^2-t_{K})^2}\,,\\
  g_{77} &=& 2(s-m_{N}^2)E_{\gamma}^{\ast}\,,\\
  g_{78} &=& 0\,,\qquad
  g_{79} = 0\,,\\
  g_{710} &=& \frac{(s-m_{N}^2)^2}{2(M_{K}^2-t_{K})}E_{\gamma}^{\ast}|\vec{q}_{K}|\sin\theta_{K}\,,\\ 
  g_{88} &=& g_{77}\,,\\
  g_{89} &=& -\frac{(s-m_{N}^2)^2}{2(M_{K}^2-t_{K})}E_{\gamma}^{\ast}|\vec{q}_{K}|\sin\theta_{K}\,,\qquad
  g_{810} = 0\,,\\
  g_{99} &=& \frac{(s-m_{N}^2)^3}{(M_{K}^2-t_{K})^2}E_{\gamma}^{\ast}|\vec{q}_{K}|^2\sin^2\theta_{K}\,,\qquad\,
  g_{910} = 0\,,\qquad
  g_{1010} = g_{99}\,.
\end{eqnarray*}
For the special form of the pole terms in Eq.~(\ref{eq:C1p_poleapprox}), the formula for $|\mathcal{A}|^2_{1+}$ reduces to Eq.~(\ref{eq:Asqr_poleapprox}).

\newpage

\section{Invariant amplitudes from $\Sigma^{\ast 0}$ exchange graphs}
\label{app:Mi_decuplet}
\def\theequation{\Alph{section}.\arabic{equation}}
\setcounter{equation}{0}

We split the computation of the $\Sigma^{\ast 0}$ exchange graphs in two parts, consisting of the graphs with the three possible photon insertions on the l.h.s. (part ($a$)) and on the r.h.s (part ($b$)) of the decuplet propagator in Fig.~\ref{fig:Sigstar0exchange}\,. Only part ($a$) contains a pole at $M_{\pi\Sigma}=m_{D}=m_{\Sigma^{\ast 0}}$\,. 
\begin{eqnarray*}
  \mathcal{M}_{1}^{(a)} &=& \frac{e\,\mathcal{C}^2}{36F_{\pi}F_{K}}\frac{\nu_{\pi\Sigma}\,Q_{K}}{m_{D}^2-M_{\pi\Sigma}^2}\biggl[3(q_{\pi}\cdot q_{K})-2\alpha^2E_{\pi}^{\ast}E_{K}^{\ast} - (m_{\Sigma}+m_{D})\left(\alpha(E_{\pi}^{\ast} + E_{K}^{\ast})+m_{N}+m_{\Sigma}\right) \\ & & \hspace{4cm} +\,\,M_{\pi}^2+(m_{N}+m_{\Sigma})\alpha E_{\pi}^{\ast}\biggr]\,,\\
  \mathcal{M}_{2}^{(a)} &=& \frac{e\,\mathcal{C}^2}{36F_{\pi}F_{K}}\frac{\nu_{\pi\Sigma}}{m_{D}^2-M_{\pi\Sigma}^2}\biggl[\left(\frac{2Q_{N}}{s-m_{N}^2}\right)\biggl((m_{N}+m_{D})\left(3(q_{\pi}\cdot q_{K})-2\alpha^2E_{\pi}^{\ast}E_{K}^{\ast}\right) - M_{\pi}^2\alpha E_{K}^{\ast} \\ & & \hspace{3.7cm} +\,\,(s-M_{\pi\Sigma}^2)\left(\alpha E_{\pi}^{\ast}-(m_{\Sigma}+m_{D})\right) + (m_{\Sigma}+m_{D})(m_{\Sigma}-m_{N})\alpha E_{K}^{\ast}\biggr) \\ & & \hspace{1.7cm} -\,\,\frac{Q_{K}}{m_{D}}\biggl((m_{\Sigma}+m_{D})\left(m_{\Sigma}+m_{D}-m_{N}\right)-3m_{D}(m_{\Sigma}-m_{N})-M_{\pi}^2 - 2m_{N}\alpha E_{\pi}^{\ast}\biggr)\biggr]\,,\\
  \mathcal{M}_{3}^{(a)} &=& \frac{e\,\mathcal{C}^2}{36F_{\pi}F_{K}}\frac{3\nu_{\pi\Sigma}\,Q_{K}}{m_{D}^2-M_{\pi\Sigma}^2}(m_{N}+m_{D})\,,\\
  \mathcal{M}_{4}^{(a)} &=& \frac{e\,\mathcal{C}^2}{36F_{\pi}F_{K}}\frac{\nu_{\pi\Sigma}}{m_{D}^2-M_{\pi\Sigma}^2}\biggl[\left(\frac{2Q_{K}}{t_{K}-M_{K}^2}\right)\biggl((m_{N}+m_{D})\left(3(q_{\pi}\cdot q_{K})-2\alpha^2E_{\pi}^{\ast}(E_{K}^{\ast}-E_{\gamma}^{\ast})\right) \\ & & \hspace{2.5cm} +\,\,\frac{3}{2}(m_{N}+m_{D})(t_{\pi}-M_{\pi}^2) - M_{\pi}^2\alpha(E_{K}^{\ast}-E_{\gamma}^{\ast}) -\alpha E_{\pi}^{\ast}(M_{\pi\Sigma}^2-m_{N}^2)  \\ & & \hspace{4cm} +\,\,(m_{\Sigma}+m_{D})\left(M_{\pi\Sigma}^2-m_{N}^2+(m_{\Sigma}-m_{N})\alpha(E_{K}^{\ast}-E_{\gamma}^{\ast})\right)\biggr) \\
    & & \hspace{2.5cm} +\,\,\frac{Q_{K}}{m_{D}}\biggl((m_{N}+m_{D})\left(3m_{D}-2\alpha E_{\pi}^{\ast}\right) - M_{\pi}^2+(m_{\Sigma}+m_{D})(m_{\Sigma}-m_{N})\biggr)\biggr]\,,\\
  \mathcal{M}_{5}^{(a)} &=& \frac{e\,\mathcal{C}^2}{36F_{\pi}F_{K}}\frac{\nu_{\pi\Sigma}}{m_{D}^2-M_{\pi\Sigma}^2}\biggl[\left(\frac{Q_{N}}{s-m_{N}^2}\right)\biggl((m_{N}+m_{D})\left(3(q_{\pi}\cdot q_{K})-2\alpha^2E_{\pi}^{\ast}E_{K}^{\ast}\right) - M_{\pi}^2\alpha E_{K}^{\ast} \\ & & \hspace{4cm} +\,\,(s-M_{\pi\Sigma}^2)\left(\alpha E_{\pi}^{\ast}-(m_{\Sigma}+m_{D})\right) + (m_{\Sigma}+m_{D})(m_{\Sigma}-m_{N})\alpha E_{K}^{\ast}\biggr) \\ & & \hspace{4cm} +\,\,Q_{K}\left(m_{\Sigma}+m_{D} - \alpha E_{\pi}^{\ast}\right) \biggr]\,,\\
  \mathcal{M}_{6}^{(a)} &=& \frac{e\,\mathcal{C}^2}{36F_{\pi}F_{K}}\frac{\nu_{\pi\Sigma}}{m_{D}^2-M_{\pi\Sigma}^2}\left(-\frac{Q_{K}}{m_{D}}\right)\left(2m_{D} - m_{\Sigma} - 2\alpha E_{\pi}^{\ast}\right)\,,\\
  \mathcal{M}_{7}^{(a)} &=& \frac{e\,\mathcal{C}^2}{36F_{\pi}F_{K}}\frac{3\nu_{\pi\Sigma}\,Q_{K}}{m_{D}^2-M_{\pi\Sigma}^2}\,,\\
  \mathcal{M}_{8}^{(a)} &=& \frac{e\,\mathcal{C}^2}{36F_{\pi}F_{K}}\frac{\nu_{\pi\Sigma}}{m_{D}^2-M_{\pi\Sigma}^2}\biggl[\left(\frac{2Q_{K}}{t_{K}-M_{K}^2}\right)\biggl(3(q_{\pi}\cdot q_{K})-2\alpha^2E_{\pi}^{\ast}(E_{K}^{\ast}-E_{\gamma}^{\ast}) + \frac{3}{2}t_{\pi}-\frac{1}{2}M_{\pi}^2\\ & & \hspace{3.5cm} -\,\,(m_{\Sigma}+m_{D})(\alpha(E_{K}^{\ast}-E_{\gamma}^{\ast}) + m_{N}+m_{\Sigma}) - (m_{D}-m_{N})\alpha E_{\pi}^{\ast}\biggr)  \\ & & \hspace{9.25cm} +\,\,\frac{Q_{K}}{m_{D}}\left(2m_{D} - m_{\Sigma} - 2\alpha E_{\pi}^{\ast}\right)\biggr]\,,\\
  \mathcal{M}_{9}^{(a)} &=& 0\,,\qquad 
  \mathcal{M}_{10}^{(a)} = -\left(\frac{2\mathcal{M}_{1}^{(a)}}{s-m_{N}^2} + \mathcal{M}_{6}^{(a)}\right)\,,\\
  \mathcal{M}_{11}^{(a)} &=& -\mathcal{M}_{7}^{(a)}\,,\quad \mathcal{M}_{12}^{(a)} = -\mathcal{M}_{8}^{(a)}\,,\quad \mathcal{M}_{13}^{(a)} = -\frac{\mathcal{M}_{1}^{(a)}}{s-m_{N}^2}\,,\quad 
  \mathcal{M}_{14}^{(a)} = \mathcal{M}_{15}^{(a)} = \mathcal{M}_{16}^{(a)} = 0\,.
\end{eqnarray*}
Here we have used the abbreviations
\begin{displaymath}
\alpha := \frac{M_{\pi\Sigma}}{m_{D}}\,,\quad (q_{\pi}\cdot q_{K}) = \frac{1}{2}\left(s+t_{\Sigma}+u_{\Sigma} - m_{\Sigma}^2 - m_{N}^2 - M_{\pi}^2 - M_{K}^2\right)\,,
\end{displaymath}
and we recall the following quantities defined in \cite{Bruns:2020lyb} and App.~\ref{app:energies_momenta},
\begin{eqnarray*}
  t_{\pi} &=& M_{\pi}^2 -\left(s-m_{N}^2 + u_{\Sigma}-m_{\Sigma}^2 + t_{K} - M_{K}^2\right)\,,\\
  E_{\pi}^{\ast} &=& M_{\pi\Sigma} - E_{\Sigma}^{\ast} = \frac{1}{2M_{\pi\Sigma}}\left(M_{\pi\Sigma}^2 + M_{\pi}^2 - m_{\Sigma}^2\right)\,,\quad E_{K}^{\ast} =  \frac{1}{2M_{\pi\Sigma}}\left(s - M_{K}^2 - M_{\pi\Sigma}^2\right)\,,\\
  E_{\gamma}^{\ast} &=& \frac{1}{2M_{\pi\Sigma}}\left(s-m_{N}^2 + t_{K} - M_{K}^2\right)\,, \quad E_{N}^{\ast} = \frac{1}{2M_{\pi\Sigma}}\left(M_{\pi\Sigma}^2 + m_{N}^2 - t_{K}\right)\,.
\end{eqnarray*}
With these expressions, and using that the nucleon charge $Q_{N}=Q_{K}$ in the case considered here, one can verify that already the above subclass of amplitudes is gauge-invariant for itself (since the exchanged decuplet member is neutral). For subclass ($b$), we can use $Q_{\pi}+Q_{\Sigma}=0$\,:
\begin{eqnarray*}
  \mathcal{M}_{1}^{(b)} &=& \frac{e\,\mathcal{C}^2}{36F_{\pi}F_{K}}\frac{\nu_{\pi\Sigma}\,(t_{K}-M_{K}^2)}{m_{D}^2-(p_{N}-q_{K})^2}\left(\frac{Q_{\Sigma}}{u_{\Sigma}-m_{\Sigma}^2}\right)\biggl[(m_{\Sigma}+m_{D})\left(m_{N}+m_{\Sigma} + \alpha\tilde{E}_{K}^{\ast}\right) \\ & & \hspace{2.5cm} +\,\, (m_{D}-m_{N})\alpha\tilde{E}_{\pi}^{\ast} + u_{\Sigma}-m_{\Sigma}^2 -M_{\pi}^2 -\left(3(q_{\pi}\cdot q_{K})-2\alpha^2\tilde{E}_{\pi}^{\ast}\tilde{E}_{K}^{\ast}\right)\biggr]\,,\\
  \mathcal{M}_{2}^{(b)} &=& \frac{e\,\mathcal{C}^2}{36F_{\pi}F_{K}}\frac{\nu_{\pi\Sigma}}{m_{D}^2-(p_{N}-q_{K})^2}\left(\frac{2Q_{\pi}}{t_{\pi}-M_{\pi}^2}\right)\biggl[(m_{N}+m_{D})\left(3(q_{\pi}\cdot q_{K})-2\alpha^2(\tilde{E}_{\pi}^{\ast}-E_{\gamma}^{\ast})\tilde{E}_{K}^{\ast}\right) \\  & & \hspace{2.5cm} +\,\,\frac{3}{2}(m_{N}+m_{D})(t_{K}-M_{K}^2) + \alpha(\tilde{E}_{\pi}^{\ast}-E_{\gamma}^{\ast})\left(2m_{D}\alpha\tilde{E}_{K}^{\ast}+M_{K}^2\right) \\  & & \hspace{2.5cm} +\,\, \alpha\tilde{E}_{K}^{\ast}\left(\frac{m_{N}}{m_{D}}(t_{\pi}-M_{\pi}^2) -M_{\pi}^2 -(m_{\Sigma}+m_{D})(2m_{D}+m_{N}-m_{\Sigma})\right) \\  & & \hspace{8.2cm} -\,\,\frac{M_{K}^2}{2m_{D}}\left(t_{\pi}-M_{\pi}^2 +2m_{D}(m_{\Sigma}+m_{D})\right)\biggr]\,,\\
  \mathcal{M}_{3}^{(b)} &=& \frac{e\,\mathcal{C}^2}{36F_{\pi}F_{K}}\frac{\nu_{\pi\Sigma}}{m_{D}^2-(p_{N}-q_{K})^2}\biggl[\left(\frac{2Q_{\Sigma}}{u_{\Sigma}-m_{\Sigma}^2}\right)\biggl(M_{K}^2\left(\alpha\tilde{E}_{\pi}^{\ast}-(m_{\Sigma}+m_{D})\right) \\ & & \hspace{2.5cm} +\,\, \alpha\tilde{E}_{K}^{\ast}\bigl(2m_{D}\alpha\tilde{E}_{\pi}^{\ast} + u_{\Sigma}-m_{\Sigma}^2 -M_{\pi}^2 -(m_{\Sigma}+m_{D})(2m_{D}+m_{N}-m_{\Sigma})\bigr) \\ & & \hspace{7.7cm} +\,\, (m_{N}+m_{D})\left(3(q_{\pi}\cdot q_{K})-2\alpha^2\tilde{E}_{\pi}^{\ast}\tilde{E}_{K}^{\ast}\right)\biggr) \\ & & \hspace{2.5cm} -\,\,\left(\frac{2Q_{\pi}}{t_{\pi}-M_{\pi}^2}\right)\biggl((m_{N}+m_{D})\left(3(q_{\pi}\cdot q_{K})-2\alpha^2(\tilde{E}_{\pi}^{\ast}-E_{\gamma}^{\ast})\tilde{E}_{K}^{\ast}\right) \\  & & \hspace{3.5cm} +\,\,\frac{3}{2}(m_{N}+m_{D})(t_{K}-M_{K}^2) + \alpha(\tilde{E}_{\pi}^{\ast}-E_{\gamma}^{\ast})\left(2m_{D}\alpha\tilde{E}_{K}^{\ast}+M_{K}^2\right) \\  & & \hspace{3.5cm} -\,\, \alpha\tilde{E}_{K}^{\ast}\left((m_{\Sigma}+m_{D})(2m_{D}+m_{N}-m_{\Sigma}) + t_{\pi}\right) - M_{K}^2(m_{\Sigma}+m_{D})\biggr)\biggr]\,,\\
  \mathcal{M}_{4}^{(b)} &=& \frac{e\,\mathcal{C}^2}{36F_{\pi}F_{K}}\frac{\nu_{\pi\Sigma}}{m_{D}^2-(p_{N}-q_{K})^2}\biggl[\frac{Q_{\Sigma}}{m_{D}}\biggl((m_{N}+m_{D})\left(3m_{D}+2\alpha\tilde{E}_{K}^{\ast}\right) - M_{K}^2 \biggr) \\ & & \hspace{2.5cm} -\,\,\left(\frac{2Q_{\pi}}{t_{\pi}-M_{\pi}^2}\right)\biggl((m_{N}+m_{D})\left(3(q_{\pi}\cdot q_{K})-2\alpha^2(\tilde{E}_{\pi}^{\ast}-E_{\gamma}^{\ast})\tilde{E}_{K}^{\ast}\right) \\  & & \hspace{3.1cm} +\,\,\frac{3}{2}(m_{N}+m_{D})(t_{K}-M_{K}^2) + \alpha(\tilde{E}_{\pi}^{\ast}-E_{\gamma}^{\ast})\left(2m_{D}\alpha\tilde{E}_{K}^{\ast}+M_{K}^2\right) \\  & & \hspace{3.1cm} -\,\, \alpha\tilde{E}_{K}^{\ast}\left((m_{\Sigma}+m_{D})(2m_{D}+m_{N}-m_{\Sigma}) + t_{\pi}\right) - M_{K}^2(m_{\Sigma}+m_{D})\biggr)\biggr]\,,\\
  \mathcal{M}_{5}^{(b)} &=& \frac{e\,\mathcal{C}^2}{36F_{\pi}F_{K}}\frac{\nu_{\pi\Sigma}}{m_{D}^2-(p_{N}-q_{K})^2}\left(\frac{Q_{\Sigma}}{u_{\Sigma}-m_{\Sigma}^2}\right)\biggl[M_{K}^2\left(\alpha\tilde{E}_{\pi}^{\ast}-(m_{\Sigma}+m_{D})\right) \\ & & \hspace{2.5cm} +\,\, \alpha\tilde{E}_{K}^{\ast}\bigl(2m_{D}\alpha\tilde{E}_{\pi}^{\ast} + u_{\Sigma}-m_{\Sigma}^2 -M_{\pi}^2 -(m_{\Sigma}+m_{D})(2m_{D}+m_{N}-m_{\Sigma})\bigr) \\ & & \hspace{7.7cm} +\,\, (m_{N}+m_{D})\left(3(q_{\pi}\cdot q_{K})-2\alpha^2\tilde{E}_{\pi}^{\ast}\tilde{E}_{K}^{\ast}\right)\biggr]\,,\\
  \mathcal{M}_{6}^{(b)} &=& \mathcal{M}_{7}^{(b)} =0\,,\quad  \mathcal{M}_{8}^{(b)} = \frac{2\mathcal{M}_{1}^{(b)}}{t_{K}-M_{K}^2}\,,\quad \mathcal{M}_{9}^{(b)} =0\,,\\
  \mathcal{M}_{10}^{(b)} &=& \frac{e\,\mathcal{C}^2}{36F_{\pi}F_{K}}\frac{\nu_{\pi\Sigma}}{m_{D}^2-(p_{N}-q_{K})^2}\left(\frac{2Q_{\pi}}{t_{\pi}-M_{\pi}^2}\right)\biggl[-\left(3(q_{\pi}\cdot q_{K})-2\alpha^2(\tilde{E}_{\pi}^{\ast}-E_{\gamma}^{\ast})\tilde{E}_{K}^{\ast}\right) -M_{\pi}^2 \\  & & \hspace{2.5cm} +\,\,\frac{3}{2}(M_{K}^2-t_{K}) + (m_{\Sigma}+m_{D})(m_{N}+m_{\Sigma}) + \alpha(\tilde{E}_{\pi}^{\ast}-E_{\gamma}^{\ast})(m_{D}-m_{N}) \\  & & \hspace{2.5cm} +\,\,  \frac{(m_{N}-m_{D})}{2m_{D}}(t_{\pi}-M_{\pi}^2) + \frac{\alpha\tilde{E}_{K}^{\ast}}{m_{D}}\left(m_{D}(m_{\Sigma}+m_{D}) + M_{\pi}^2-t_{\pi}\right)\biggr]\,, \\
  \mathcal{M}_{11}^{(b)} &=& \frac{e\,\mathcal{C}^2}{36F_{\pi}F_{K}}\frac{\nu_{\pi\Sigma}}{m_{D}^2-(p_{N}-q_{K})^2}\biggl[\left(\frac{2Q_{\Sigma}}{u_{\Sigma}-m_{\Sigma}^2}\right)\biggl((m_{\Sigma}+m_{D})\left(m_{N}+m_{\Sigma} + \alpha\tilde{E}_{K}^{\ast}\right) \\ & & \hspace{2.5cm} +\,\, (m_{D}-m_{N})\alpha\tilde{E}_{\pi}^{\ast} + u_{\Sigma}-m_{\Sigma}^2 -M_{\pi}^2 -\left(3(q_{\pi}\cdot q_{K})-2\alpha^2\tilde{E}_{\pi}^{\ast}\tilde{E}_{K}^{\ast}\right)\biggr) \\  & & \hspace{2.5cm} +\,\,\left(\frac{2Q_{\pi}}{t_{\pi}-M_{\pi}^2}\right)\biggl(3(q_{\pi}\cdot q_{K})-2\alpha^2(\tilde{E}_{\pi}^{\ast}-E_{\gamma}^{\ast})\tilde{E}_{K}^{\ast} + \frac{3}{2}(t_{K}-M_{K}^2) \\  & & \hspace{2.5cm} +\,\,t_{\pi} - (m_{\Sigma}+m_{D})(m_{N}+m_{\Sigma} + \alpha\tilde{E}_{K}^{\ast}) - \alpha(\tilde{E}_{\pi}^{\ast}-E_{\gamma}^{\ast})(m_{D}-m_{N})\biggr)\biggr]\,,\\
  \mathcal{M}_{12}^{(b)} &=& \frac{e\,\mathcal{C}^2}{36F_{\pi}F_{K}}\frac{\nu_{\pi\Sigma}}{m_{D}^2-(p_{N}-q_{K})^2}\biggl[\frac{Q_{\pi}}{m_{D}}\biggl(2m_{D}-m_{N} + 2\alpha\tilde{E}_{K}^{\ast} \biggr) \\  & & \hspace{2.5cm} +\,\,\left(\frac{2Q_{\pi}}{t_{\pi}-M_{\pi}^2}\right)\biggl(3(q_{\pi}\cdot q_{K})-2\alpha^2(\tilde{E}_{\pi}^{\ast}-E_{\gamma}^{\ast})\tilde{E}_{K}^{\ast} + \frac{3}{2}(t_{K}-M_{K}^2) \\  & & \hspace{2.5cm} +\,\,t_{\pi} - (m_{\Sigma}+m_{D})(m_{N}+m_{\Sigma} + \alpha\tilde{E}_{K}^{\ast}) - \alpha(\tilde{E}_{\pi}^{\ast}-E_{\gamma}^{\ast})(m_{D}-m_{N})\biggr)\biggr]\,,\\
  \mathcal{M}_{13}^{(b)} &=& \frac{1}{2} \mathcal{M}_{8}^{(b)}\,,\quad \mathcal{M}_{14}^{(b)} = \mathcal{M}_{15}^{(b)} = \mathcal{M}_{16}^{(b)} = 0\,, \,\,\,\quad (p_{N}-q_{K})^2=M_{\pi\Sigma}^2+m_{N}^2-s+M_{K}^2-t_{K}\,.
\end{eqnarray*}
For our convenience, we have used the new abbreviations
\begin{equation*}
\tilde{E}_{\pi}^{\ast} := E_{\pi}^{\ast} + \frac{t_{\pi}-M_{\pi}^2}{2M_{\pi\Sigma}}\,,\qquad \tilde{E}_{K}^{\ast} := E_{K}^{\ast} + \frac{t_{K}-M_{K}^2}{2M_{\pi\Sigma}}\,.
\end{equation*}

\newpage

\end{appendix}

\end{document}